\documentclass[journal]{IEEEtran}
\usepackage{algorithm}
\usepackage{algorithmic}
\usepackage{array}

\newtheorem{myProp}{\textit{Proposition}}

\newtheorem{myRemark}{\textit{Remark}}
\usepackage{stfloats}
\usepackage{cite}
\usepackage{amsmath,amssymb,amsfonts,bm}
\allowdisplaybreaks
\usepackage{graphicx}
\usepackage{epstopdf}
\usepackage[caption=false, font=footnotesize, labelfont=rm, textfont=rm]{subfig}
\newcommand{\subsubsubsection}[1]{\paragraph{#1}}
\usepackage{textcomp}
\usepackage{stfloats}
\usepackage{url}
\usepackage{verbatim}
\usepackage{enumerate}
\usepackage{amssymb}
\usepackage{xcolor}
\def\BibTeX{{\rm B\kern-.05em{\sc i\kern-.025em b}\kern-.08em
		T\kern-.1667em\lower.7ex\hbox{E}\kern-.125emX}}
\usepackage{xpatch}
\makeatletter
\def\changeBibColor#1{%
	\in@{#1}{ }
	\ifin@\color{blue}\else\normalcolor\fi
}
\xpatchcmd
\@bibitem
{\item}
{\changeBibColor{#1}\item}
{}{\fail}
\xpatchcmd\@lbibitem
{\item}
{\changeBibColor{#2}\item}
{}{\fail}
\makeatother

\usepackage{hyperref}
\hypersetup{colorlinks=true, linkcolor=red, citecolor=blue, filecolor=blue, urlcolor=black}

\begin{document}

\title{Pinching-Antenna Systems (PASS)-Enabled \\ UAV Delivery}

\author{Suyu Lv, Meng Li,~\IEEEmembership{Senior Member,~IEEE}, Qi Li,~\IEEEmembership{Member,~IEEE} and Yuanwei Liu,~\IEEEmembership{Fellow,~IEEE}
	\thanks{Suyu Lv, Meng Li and Qi li are with School of Information Science and Technology, Beijing University of Technology, Beijing, 100124, China (e-mail: lvsuyu@bjut.edu.cn, limeng720@bjut.edu.cn, liqi0301@bjut.edu.cn).}
	\thanks{Yuanwei Liu is with the Department of Electrical and Electronic Engineering, the University of Hong Kong, Hong Kong (e-mail: yuanwei@hku.hk).}
}

\maketitle

\begin{abstract}
	A pinching-antenna systems (PASS)-enabled unmanned aerial vehicle (UAV) delivery framework is proposed, which exploits the capability of PASS to establish a strong line-of-sight link and reduce free-space pathloss.
	Aiming at minimizing the communication energy consumption in one cycle, a double-layer optimization (DLO) algorithm is developed by jointly optimizing the UAV delivery sequence and the pinching antenna (PA) activation vector. 
	More specifically, {\textit{at the outer layer}}, a hierarchical alternating optimization (HAO) scheme is proposed to tackle the NP-hard problem of delivery sequence planning, where a genetic algorithm performs global exploration to generate candidate solutions at the top-level, while a dynamic programming performs local refinement to obtain elite solutions at the lower-level. 
	With determined UAV trajectory, {\textit{at the inner layer}}, focus is placed on addressing the highly coupled mixed-integer nonlinear programming problem of PA activation vector optimization, where a pair of algorithms are proposed: 1) Branch-and-Bound (BnB) algorithm for finding global optimum; 2) incremental search and local refinement (ISLR) algorithm for reducing computational complexity. 
	Simulation results indicate that: i) The proposed HAO-based delivery sequence planning scheme can effectively reduce the total flight distance, thereby decreasing flight time and communication energy consumption; ii) Both the proposed BnB and ISLR algorithms can achieve energy-efficient PA activation, with the former exhibiting better performance and the latter having lower complexity; iii) PASS outperforms the conventional multi-antenna systems, especially with higher communication rate requirements.
\end{abstract}

\begin{IEEEkeywords}
	Antenna activation, delivery sequence planning,  pinching-antenna systems (PASS), unmanned aerial vehicle (UAV).
\end{IEEEkeywords}

\section{Introduction}

In recent decades, unmanned aerial vehicle (UAV) technology has undergone unprecedented growth, transitioning from military applications to civilian domains such as logistics, agriculture, emergency rescue, and environmental monitoring \cite{COMST.2019.2902862, ACCESS.2019.2909530}. 
Among these, UAV delivery stands out as a transformative solution to address inefficiencies in traditional logistics, particularly in last-mile delivery, remote area supply, and post-disaster material transportation \cite{TNSM.2024.3487357}. 
The reliability of UAV delivery systems is heavily dependent on stable real-time communication with ground base stations (BS) for trajectory adjustments, obstacle avoidance signals, and delivery coordinates \cite{TGCN.2023.3273951}. 
However, UAV mobility causes dynamic communication distances, leading to significant free-space pathloss and high energy consumption in conventional multi-antenna systems, which struggle to balance efficiency and reliability under high-rate requirements.

To address the severe free-space pathloss, which is a problem aggravated in high-frequency UAV communications, researchers have explored novel flexible antenna technologies as alternatives to conventional multi-input-multi-output (MIMO) systems. 
For instance, reconfigurable intelligent surfaces (RIS) can actively reconfigure the wireless propagation environment by adjusting meta-element phase shifts, yet they suffer from double-fading issues \cite{JSAC.2020.3007211, LWC.2022.3221609}. 
Fluid antennas leverage conductive fluid deformability to optimize radiation patterns, but their stability is compromised by unstable shape and temperature sensitivity \cite{TWC.2023.3276245, OJAP.2021.3069325}. 
Movable antennas are restricted to a movement range of just a few wavelengths, limiting adaptability to long-distance UAV flights \cite{MCOM.001.2300212}. 
These inherent limitations prevent these technologies from providing flexible, cost-effective, and practically deployable support for dynamic UAV delivery communication, highlighting the need for a more feasible antenna solution. 

Pinching-antenna systems (PASS) represent a groundbreaking advancement in flexible antenna technology, first demonstrated by NTT DOCOMO \cite{DOCOMO, DOCOMO_PDF}. 
PASS comprises dielectric waveguides serving as low-attenuation signal carriers, and pinching antennas (PAs) serving as contactless directional couplers that extract energy from waveguides and radiate signals toward the free space. 
Due to the simple yet scalable structure, PASS enables meter-scale PA position adjustment along the waveguide.
By activating PAs at locations close to the UAV, it shortens the free-space propagation distance, drastically reducing pathloss while creating stable line-of-sight (LoS) links \cite{TCOMM.2025.3555866, MCOM.001.2500037}. 
This unique capability allows PASS to ``pass" signals to UAVs from near-UAV PAs, and leverages near-field channel advantages from its large waveguide aperture, effectively shifting wireless communication from the ``last mile" to the ``last meter" paradigm \cite{arXiv:2508.07572}. 
Additionally, another notable advantage of PASS is its energy efficiency, which stems from the extremely low-attenuation in-waveguide transmission, the short-distance free-space propagation, and the selective PA activation mechanism, avoiding unnecessary energy waste. 
These traits make PASS suitable for tackling the challenges of UAV delivery systems, where stable communication and energy efficiency are of crucial importance. 

Prior studies on PASS have validated its potential in static scenarios. 
The authors in \cite{LCOMM.2025.3566299} quantified how PASS array gain correlates with antenna spacing, proving there exists an optimal spacing to maximize array gain, which laid a critical theoretical foundation for antenna spacing design in PASS. 
Building on this, the authors in \cite{arXiv:2506.14298} focused on single-waveguide multi-user scenarios, investigating how pinching beamforming shapes capacity regions and achievable rate regions.
Extending to emerging application scenarios, the authors in \cite{arXiv:2505.10179} expanded PASS performance analysis to integrated sensing and communication (ISAC), showing that PASS delivers a larger rate region for both communication and sensing compared to traditional antenna-enabled ISAC systems. 
For multiple access selection, the authors in \cite{arXiv:2508.05309} compared non-orthogonal multiple access (NOMA) and orthogonal multiple access (OMA) in PASS, revealing that OMA outperforms NOMA in sum capacity when pinching beamforming vectors are large, while also simplifying transceiver structures. 
The authors in \cite{TCCN.2025.3564470} provided a universal evaluation tool by deriving closed-form expressions of key metrics under both ideal and practical waveguide conditions. 
LoS blockage impact on PASS was analyzed in \cite{LWC.2025.3579616}, showing PASS's enclosed waveguide transmission mitigates external interference, validating its suitability for complex blocked environments. 
The authors in \cite{arXiv:2506.05102} compared PASS with RIS in millimeter-wave bands, noting RIS requires ultra-large-scale configuration to match PASS's performance, and PASS is more robust to hardware damage or severe pathloss, confirming its better adaptability to high-frequency communication.
These theoretical studies on the achievable performance of PASS have established an analytical framework, confirming the superiority of PASS over traditional MIMO from multiple perspectives. 

Research on PASS resource scheduling and performance optimization centers on PA placement, which is a core variable that directly determines pathloss, coupling efficiency and array gain, thus governing overall transmission performance. 
The authors in \cite{arXiv:2507.13307} studied an analytical optimization model for PASS antenna placement, clearly quantifying the correlation between PA position and system performance, and deriving closed-form solutions for optimal PA locations. 
To bridge theory and practice, the authors in \cite{LCOMM.2025.3574633} proposed a low-complexity PA placement scheme, addressing the high computational costs of ideal optimization models and enabling real-world deployment. 
For multi-user multicast downlink links, the authors in \cite{OJCOMS.2025.3582895} explored dynamic PA placement to enhance resource efficiency, considering both continuous and discrete PA activation modes to strike a balance between performance and complexity for practical use cases. 
Focusing on downlink single-user rate maximization, the authors in \cite{LWC.2025.3543889} adopted a two-stage approach, first optimizing PA positions to minimize large-scale pathloss, then refining positions to boost received signal strength, thus achieving a practical trade-off between optimization accuracy and complexity. 
Shifting to uplink scenarios, the authors in \cite{LWC.2025.3547956} targeted maximizing the minimum data rate to ensure user fairness and suppress interference, verifying PASS's advantages in combating LoS blockage and large-scale fading and filling the gap in uplink fairness optimization. 
For single-waveguide systems, NOMA is a natural choice for multi-user service to overcome the limitations of single data stream. 
Following this thought, the authors in \cite{LWC.2025.3578312} and \cite{LWC.2025.3548280} optimized PA activation positions to maximize sum rate in downlink NOMA PASS, while the authors in \cite{LWC.2025.3600899} optimized PA activation positions and user transmit power to maximize energy efficiency in uplink NOMA PASS. 
Addressing quality of service (QoS) in Internet of Things, the authors in \cite{TVT.2025.3598591} optimized energy efficiency for downlink PASS under QoS constraints, highlighting PASS's energy advantages over traditional antennas. 
Extending to novel application scenarios, the authors in \cite{LWC.2025.3597719} the authors optimized PASS-aided ISAC systems by integrating PA position adjustment and user power control with a maximum entropy-based reinforcement learning algorithm, filling the gap in PASS-ISAC resource allocation. 

Nevertheless, existing studies have not integrated PASS into dynamic UAV delivery scenarios, nor addressed the critical challenge of jointly optimizing two intractable subproblems.
The first one is the UAV delivery sequence planning, which an NP-hard traveling salesman problem (TSP)-like task.
The second one is the PA activation vector optimization under dynamic environment, which is a highly coupled mixed-integer nonlinear programming (MINLP) problem.
Both subproblems are essential for minimizing communication energy while completing delivery tasks and meeting communication rate constraints, yet remain inadequately addressed in existing research. 
To fill these gaps, this article proposes a PASS-enabled UAV delivery framework, along with a double layer optimization (DLO) algorithm, to minimize communication energy within one fight cycle.
The main contributions of this article are summarized as follows. 
\begin{itemize}
	\item We propose a PASS-enabled UAV delivery system framework, where preconfigured PAs are discretely deployed on the waveguide and can be flexibly activated. To complete delivery tasks, the UAV's trajectory must cover all delivery nodes. Based on this framework, we formulate a communication energy consumption minimization problem, and develop a DLO algorithm to perform the joint optimization of the UAV delivery sequence planning and the binary PA activation vector, while satisfying the minimum communication rate constraint. 
	\item To solve the NP-hard TSP-like problem of delivery sequence planning, we develop a hierarchical alternating optimization (HAO) scheme, which balances solution diversity and accuracy, effectively reducing UAV flight distance and time.  
	More specifically, a genetic algorithm (GA)-based scheme performs coarse-grained global exploration to generate candidate solutions {\textit{at the top level}}, and a dynamic programming (DP)-based scheme conducts fine-grained local refinement to obtain elite solutions {\textit{at the lower level}}. 
	\item With optimized UAV trajectory, we design a pair of algorithms for the MINLP problem of PA activation vector optimization, where one is for optimum and the other one is for low complexity. 
	Specifically, first, we propose an branch-and-bound (BnB)-based algorithm to find the global optimal solutions. 
	Subsequently, to further reduce the computational complexity, we develop a incremental search and local refinement (ISLR)-based algorithm to obtain sub-optimal solutions.  
	\item The effectiveness of the proposed framework and algorithms is validated via simulations, which show that i) The HAO-based delivery sequence scheme reduces total flight distance, thereby lowering flight time and communication energy consumption; ii) Both BnB and ISLR achieve energy-efficient PA activation, where BnB outperforms in energy consumption and ISLR outperforms in complexity; iii) PASS outperforms conventional MIMO systems, especially under high communication rate requirements. 
\end{itemize}

The rest of this article follows a clear structure. Section {\ref{SystemModel}} introduces the PASS-enabled UAV delivery system model and formulates the communication energy minimization problem. Section {\ref{Solution}} proposes a DLO solution, where HAO handles delivery sequence planning, with BnB (optimal) and ISLR (low-complexity) for PA activation. Section {\ref{Simulation}} and Section {\ref{Conclusion}} present simulation results and conclusions, respectively.

{\textit{Notations:}} $|x|$ denotes the absolute value of a real scalar $x$ and the modulus of a complex scalar. $\|\bf{x}\|$ represents the Euclidean norm of a vector $\bf{x}$. $\|\bf{X}\|$ denotes the Frobenius norm of a matrix $\bf{X}$. $\boldsymbol{1}_{N \times 1}$ and $\boldsymbol{0}_{N \times 1}$ represents an $N$-dimensional all-ones vector and all-zeros vector, respectively. ${\bf{X}}^T$ and ${\bf{X}}^H$ denote the transpose and Hermitian transpose of a matrix $\bf{X}$. $\lceil \cdot \rceil$ is the ceiling function that rounds a real number up to the nearest integer.

\section{System Model and Problem Formulation} \label{SystemModel}

\begin{figure}[t]
	\centering
	\includegraphics[width=8.5cm]{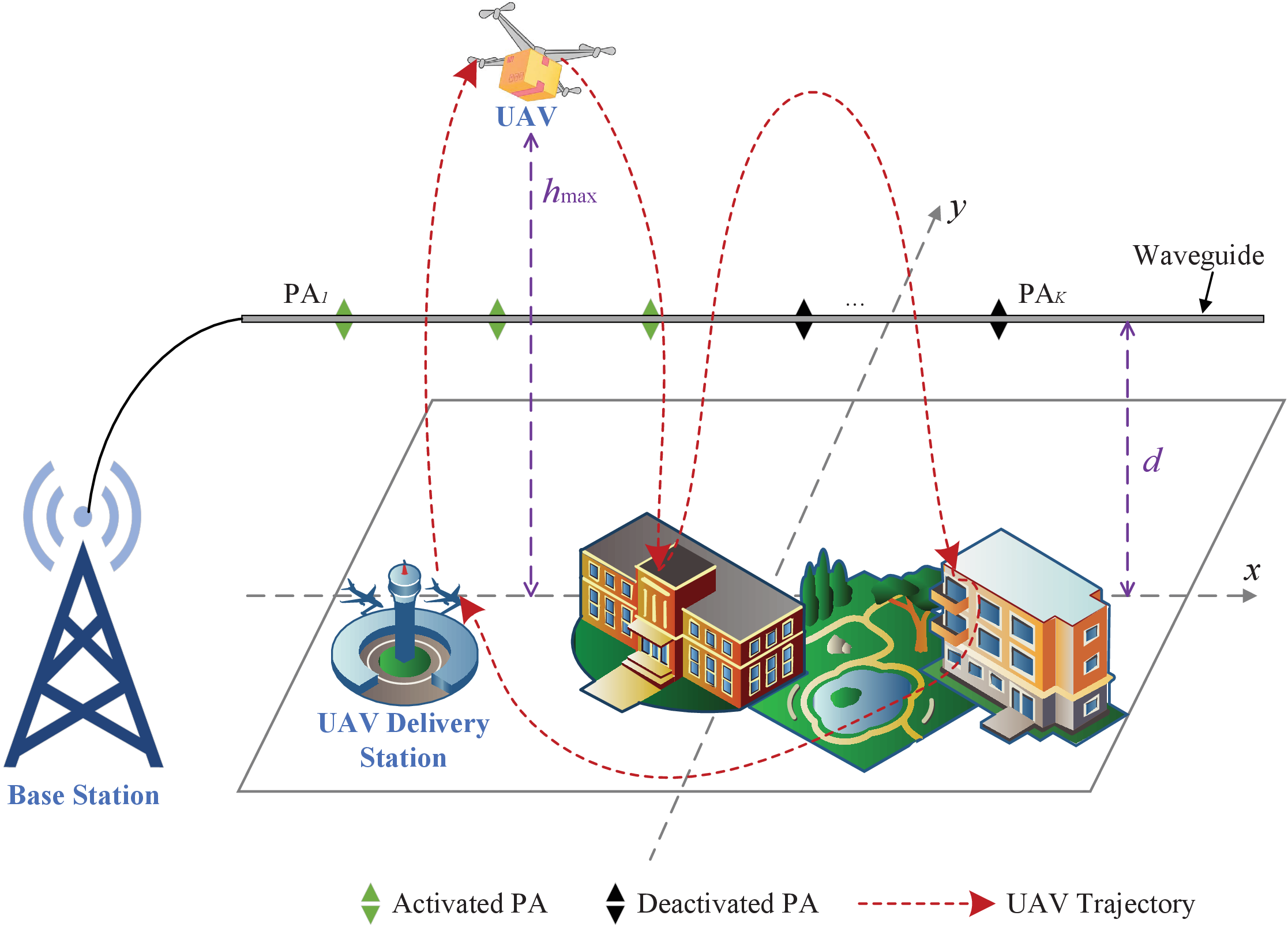}
	\caption{Illustration of PASS-enabled UAV delivery framework.}
	\label{SystemModel_PASS_UAV}
\end{figure}

In this paper, we propose a PASS-enabled UAV delivery framework, as illustrated in Fig. {\ref{SystemModel_PASS_UAV}}. 
During delivery operations, the UAV needs to receive controlling commands from the BS, thus a communication rate threshold should be met. 
The BS is equipped with a single waveguide, and $K$ available PAs are discretely distributed on the waveguide. 
The set of all available PAs is denoted by ${\cal{K}}$.

\subsection{Channel and Signal Model}

Suppose that the waveguide is placed parallel to the $x$-axis with a height of $z_{\rm{wg}}$, where the $y$-coordinate of the waveguide is $y_{\rm{wg}}$.
Denote the location of the $k$-th PA by ${\bm{\psi}}_{k}^{\rm{PA}}= \left( x_{k}^{\rm{PA}}, y_{\rm{wg}}, z_{\rm{wg}} \right) $, satisfying $\left| x_{k}^{\rm{PA}} - x_{k'}^{\rm{PA}} \right| \ge \Delta x$ to avoid mutual coupling, where $k, k' = 1, \cdots, K$ and $ k \ne k'$. 
Define a vector $\mathbf{x} = \left[ x_{1}^{\rm{PA}}, \cdots, x_{K}^{\rm{PA}} \right]^T \in {\mathbb{R}}^{K \times 1}$ to represent the $x$-coordinates of all PAs. 
Denote the location of UAV at time $t$ by ${\bm{\psi}}_{\rm{U}} \left(t\right) = \left( x_{\rm{U}} \left(t\right), y_{\rm{U}} \left(t\right), h_{\rm{U}} \left(t\right) \right)$. 
Furthermore, the Doppler effect caused by the UAV mobility is assumed to be perfectly compensated \cite{TWC.2017.2688328}.
Thus, the time-varying free-space channel between UAV and the $k$-th PA is
\begin{equation}\label{h_Unk}
	h\left(x_{k}^{\rm{PA}}, t \right)  =\frac{\sqrt{\eta}e^{- \frac{2\pi j}{\lambda } d_{{\rm{U}}, {\rm{PA}}_k}\left(t\right) }}{d_{{\rm{U}}, {\rm{PA}}_k}\left(t\right)},
\end{equation}
where $\lambda$ indicates the wavelength of the carrier frequency, $\eta  = \frac{{{c^2}}}{{16{\pi ^2}f_c^2}}$ with $c$ denoting the speed of light and $f_c$ denoting the carrier frequency, and  
$d_{{\rm{U}}, {\rm{PA}}_k}\left(t\right)$ denotes the distance between the $k$-th PA and UAV at time $t$, given by
\begin{equation}
\begin{aligned}
&  d_{{\rm{U}}, {\rm{PA}}_k}\left(t\right) 
= {\left \| {\bm{\psi}}_{\rm{U}}\left(t\right) -{\bm{\psi}}_{k}^{\rm{PA}} \right \|} \\
= & {\sqrt{\left(x_{k}^{\rm{PA}} -x_{\rm{U}} \left(t\right) \right)^2 + \left( y_{\rm{wg}} - y_{\rm{U}} \left(t\right) \right)^2 +\left(z_{\rm{wg}} - h_{\rm{U}} \left(t\right) \right)^2 }}.
\end{aligned}
\end{equation}

The free-space channel vector from the $K$ PAs to UAV is 
\begin{equation}
{\bf{h}} \left(\mathbf{x}, t \right)= \left[ h \left( x_{1}^{\rm{PA}}, t \right), \cdots, h \left( x_{K}^{\rm{PA}}, t \right) \right]^T.
\end{equation}
Therefore, the received signal at the UAV can be expressed as \begin{equation}\label{Signal_Um}
y\left(t\right) = \left[{\bf{h}}\left( \mathbf{x}, t \right)\right]^H  {\bf{g}}\left({\bm{\beta}} \left(t\right), \mathbf{x} \right) s\left(t\right) +n\left(t\right),
\end{equation}
where $s\left(t\right) $ is the signal symbol to UAV, 
$n\left(t\right) \sim \left(0, \sigma^2 \right)$ denotes the additive white Gaussian noise (AWGN) at UAV.
${\bf{g}}\left({\bm{\beta}}\left(t\right), \mathbf{x} \right) \in {\mathbb{C}} ^{K \times 1}$ represents the response vector from the feed point of the PAs  \cite{arXiv:2502.05917}. 
The $k$-the element of $\mathbf{g} \left( {\bm{\beta}}\left(t\right), \mathbf{x} \right)$ is given by
\begin{subequations}
\begin{equation}
{g} \left( \beta_k \left(t\right), {x} _{k}^{\rm{PA}} \right) = \beta_k \left(t\right) {\tilde{g}}_k,
\end{equation}
where
\begin{equation}
{\tilde{g}}_k = e^{- \frac{2\pi j}{\lambda_g } \left\| {\bm{\psi}}_{0}^{\rm{PA}} -{\bm{\psi}}_{k}^{\rm{PA}} \right\| } =  e^{- \frac{2\pi j}{\lambda_g } \left|x_{0}^{\rm{PA}} - x_{k}^{\rm{PA}} \right| },
\end{equation}
\end{subequations}
where $\lambda_g = \frac{\lambda}{n_{\rm{eff}}}$ denotes the wavelength in the waveguide with ${n_{\rm{eff}}}$ being the dielectric coefficient, ${\bm{\psi}}_{0}^{\rm{PA}} = \left( x_{0}^{\rm{PA}} , y_{\rm{wg}}, z_{\rm{wg}} \right)$ denotes the location of the feed point of the waveguide.
${\bm{\beta}} \left(t\right) = \left\{ \beta_1 \left(t\right), \cdots, \beta_K \left(t\right) \right\} \in {\mathbb{R}}^{K \times 1}$ denotes the power radiation ratio vector of PAs at time $t$, where $\beta_k \left(t\right)$ is the power radiation ratio of $k$-th PA at time $t$, satisfying $\beta_k \left(t\right) \in \left[0, 1\right]$ and ${\textstyle \sum_{k=1}^{K}} \left[\beta _k \left(t\right) \right]^2 \le 1$, and
$\beta_k \left(t\right) = 0$ means that the $k$-th PA is in the deactived state at time $t$.

\subsection{Power Radiation Model}

\begin{figure}[t]
	\centering
	\includegraphics[width=8.5cm]{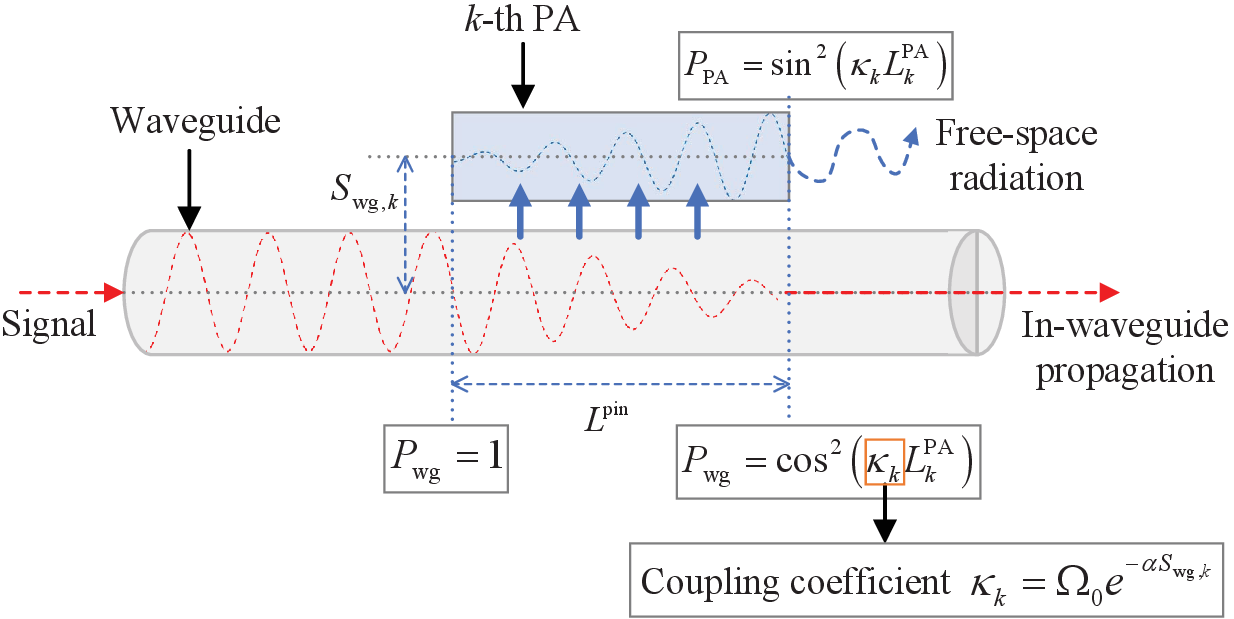}
	\caption{Illustration of power radiation model.}
	\label{Waveguide_Spacing}
\end{figure}

Expanding the analysis presented in \cite{arXiv:2502.05917}, the power radiation ratio of each PA is determined by the number and arrangement of activated PAs on the waveguide, which is expressed as
\begin{equation}
	 \beta_k\left(t\right) = a_k\left(t\right) \sin\left( \kappa_k L_k^{\rm{PA}} \right) \prod \limits_{i=1}^{k-1} \sqrt{1-a_i\left(t\right) \sin ^2 \left( \kappa_i L_i^{\rm{PA}} \right)} ,
\end{equation}
where $a_k \left(t\right) $ is the binary PA activation index, with $a_k \left(t\right) = 1$ meaning that the $k$-th PA is activated, and $a_k \left(t\right) = 0$ otherwise. $K_a \left(t\right) \triangleq \sum_{k=1}^{K} a_k \left(t\right)$ represents the number of activated PAs at time $t$.
As shown in Fig. {\ref{Waveguide_Spacing}}, $\kappa_k$ is the coupling coefficient which measures power exchange strength from the waveguide to $k$-th PA, and $L_k^{\rm{PA}}$ is the preconfigured length of $k$-th PA, which determines the coupling length. 
${P_{{\rm{wg}}}}$ and ${P_{{\rm{PA}}}}$ are the normalized radiation power of waveguide and PA, respectively. 

Based on the empirical results in \cite{TMTT.2020.3044617}, the mode coupling coefficient $\kappa_k$ can be modelled as an exponentially decaying function of spacing between the waveguide and $k$-th PA, denoted by $S_{{\rm{wg}}, k}$.
Thus, the mathematical form of $\kappa_k$ can be given by
\begin{equation}
	\kappa_k = \Omega_0 e^{-\alpha S_{{\rm{wg}}, k}},
\end{equation}
where coefficient $\Omega_0$ captures electric distribution and modal normalization, and the cladding decay constant $\alpha = \sqrt {\gamma _0^2 - \frac{{4{\pi ^2}}} {{{\lambda ^2}}} n_{{\rm{clad}}}^2}$ is determined by PAs' propagation constant $\gamma_0$ and cladding refraction index $ n_{\rm{clad}}$. 
By tuning the spacing between waveguide and PAs, the power radiation ratios can be optimally adjusted under ideal conditions \cite{arXiv:2505.00218}. 
However, in practical deployments, the length of the PAs and their spacing from the waveguide are typically predetermined. 
Therefore, to reduce hardware complexity, we assume that all PAs have the same length and equal spacing from the waveguide, i.e., $L_k^{\rm{PA}} = L^{\rm{PA}}$ and $S_{{\rm{wg}}, k} = S_{{\rm{wg}}}$, $\forall k \in {\cal{K}}$,
leading to a proportional power model
\begin{equation}\label{PowerModel}
	\beta_k\left(t\right) = a_k\left(t\right) \delta \prod \limits_{i=1}^{k-1} \sqrt{1-a_i\left(t\right) \delta^2 } ,
\end{equation}
where $ \delta = \sin\left( \Omega_0 e^{-\alpha S_{{\rm{wg}}}} L^{\rm{PA}} \right)$ can be regarded as a constant. 

The achievable communication rate of UAV is given by
\begin{equation}\label{Rate_UAV}
\begin{aligned}
& R \left( \mathbf{a} \left(t\right) , {\bm{\beta}} \left(t\right), \mathbf{x} \right) \\
= & \log _2 \left(1 + \frac{P_{\rm{C}} \left(t\right) \left| \left[ {\bf{h}} \left( \mathbf{x}, t \right)\right]^H  {\bf{G}}\left( {\bm{\beta}} \left(t\right), \mathbf{x} \right) {\mathbf{a}} \left(t\right) \right|^2} {\sigma^2} \right),
\end{aligned}
\end{equation}
where $P_{\rm{C}} \left(t\right) $ is the communication power at time $t$, $\mathbf{a} \left(t\right) = \left[ a_{1} \left(t\right), \cdots, a_{K} \left(t\right) \right]^T \in {\mathbb{Z}}^{K \times 1}$ denotes the PA activation vector, and ${\bf{G}}\left( {\bm{\beta}} \left(t\right), \mathbf{x} \right) = {\rm{diag}} \left({\bf{g}}\left( {\bm{\beta}} \left(t\right), \mathbf{x} \right)\right) $.

\subsection{UAV Delivery Model}

Suppose the number of delivery tasks is $M$, where the location of the $m$-th delivery node is denoted by ${\bf{q}}_m = \left[ x_m, y_m, h_m \right]$. 
At each delivery node, the UAV needs to hover for a certain period of time to complete the delivery tasks. 
Denote the number of delivery tasks at $m$-th delivery node by $D_{{\rm{task}}, m}$, and the delivery speed by $v_d$ (tasks per second). 
Then, the hovering time at this node is
\begin{equation}
	T_{\rm{hover}} = \sum_{m=1}^{M} \frac{D_{{\rm{task}}, m}}{v_d}.
\end{equation}

It is assumed that the UAV flies along the shortest path when traveling from delivery node ${\bf{q}}_{m}$ to ${\bf{q}}_{m+1}$. 
Moreover, for ease of analysis, we consider a constant UAV flight speed $v_f$. 
In this way, UAV trajectory planning is essentially equivalent to delivery sequence planning, which a permutation of $M$ nodes. 
Therefore, there are a total of $M!$ possible delivery sequences, denoted by ${\mathcal{E}} \triangleq \left\{ {\bf{c}}_1, \cdots, {\bf{c}}_{M!} \right\}$, where ${\bf{c}}_j, j = 1, \cdots, M!$ represents the $j$-th possible delivery sequence. 
The UAV flight time can be given by 
\begin{equation}\label{TotalTime}
	T_{\rm{flight}} = \frac{D_{\rm{Total\_path}}}{v_f},
\end{equation}
where $D_{\rm{Total\_path}}$ denotes the total distance traveled by the UAV trajectory. 

At the initial time, the UAV is located at the delivery station, with the location denoted by ${\bf{q}}_{\rm{S}}$. 
The UAV needs to return to the delivery station at the end of the flying period, i.e., 
\begin{equation}
	{\bm{\psi}}_{\rm{U}} \left(0\right) = {\bm{\psi}}_{\rm{U}} \left(T\right) = {\bf{q}}_{\rm{S}}, 
\end{equation}
where $T = T_{\rm{hover}} + T_{\rm{flight}}$ is the required time for one cycle of the UAV. 
Therefore, taking delivery sequence ${{\bf{q}}_{\rm{S}}} \to {{\bf{q}}_{1}} \to {{\bf{q}}_{2}} \to \cdots \to {{\bf{q}}_{M}} \to {{\bf{q}}_{\rm{S}}}$ as an example, the total traveled distance of UAV can be calculated by
\begin{subequations}
	\begin{equation}
		D_{\rm{Total\_path}} = \sum_{m=0}^{M} D_{{\rm{path}}, {m \to m+1}},
	\end{equation} 
	with 
	\begin{equation}
		D_{{\rm{path}}, {m \to m+1}} =
		\left\{ \begin{array}{l}
			\mathop {\left| {{{\bf{q}}_{1}} - {{\bf{q}}_{\rm{S}}}} \right|,m = 0}\limits_{}, \\
			\mathop {\left| {{{\bf{q}}_{m + 1}} - {{\bf{q}}_{m}}} \right|,m = 1, \cdots ,M - 1}\limits_{}, \\
			\left| {{{\bf{q}}_{M}} - {{\bf{q}}_{\rm{S}}}} \right|,m = M.
		\end{array} \right.
	\end{equation}
\end{subequations}

With given communication power $P_{\rm{C}} \left(t\right)$, the energy consumption of communication is
\begin{equation}\label{CommunicationEnergy}
	E_{\rm{C}} = \int_{0}^{T} P_{\rm{C}} \left(t\right) dt.
\end{equation}

Note that Eq. (\ref{CommunicationEnergy}) is formulated in continuous-time form. While this representation provides more accurate characterization, the integral in Eq. (\ref{CommunicationEnergy}) makes the problem intractable. For the sake of analysis, we consider a discretization approximation approach. 
The entire flight cycle is divided into evenly-spaced time slots, each with a duration of $\tau$. 
Thus, the required number of time slots can be calculated as
\begin{equation} \label{TimeSlots}
	L = \sum_{m=1}^{M} \lceil \frac{D_{{\rm{task}}, m}}{v_d \cdot \tau} \rceil + \sum_{m=0}^{M} \lceil \frac{D_{{\rm{path}}, {m \to m+1}}}{v_f \cdot \tau} \rceil,
\end{equation}
where $\lceil \cdot \rceil$ denotes the ceiling function.
Denote the set of time slots by ${\cal{L}} \triangleq \left\{1, \cdots, l, \cdots, L \right\}$. 
During each time slot, the UAV's location can be seen as unchanged, denoted by ${\bm{\psi}}_{{\rm{U}}} \left(l\right)$. 
Moreover, the UAV's operational mode remains unchanged within each time slot, which means that, the UAV neither switches from delivery mode to flying mode, nor switches from flying mode to delivery mode. 
Therefore, the energy consumption of communication can be reformulated as
\begin{equation}
	{{\widetilde E}_{\rm{C}}} = \sum\limits_{l = 1}^L { P_{\rm{C}} \left(l\right)\tau }. 
\end{equation}

\subsection{Problem Formulation}

The key objective of this paper is to minimize the communication energy consumption{\footnote{We consider merely the communication energy consumption, mainly due to the following reasons. 
On one hand, the flight energy consumption of the UAV and the communication energy consumption are not of the same order of magnitude. If the objective function is defined as the sum of the two, the UAV's flight power consumption will dominate the execution process of the optimization algorithm. 
On the other hand, given that the UAV's flight power is constant, minimizing the flight time is actually equivalent to optimizing the flight energy consumption.}} through joint optimization of the UAV's delivery sequence and PA activation vector, while satisfying the minimum communication rate constraint. 
The mathematical formulation of the optimization problem is 
\begin{subequations}\label{P}
\begin{align}
	({\rm{P}})\; & \mathop {\min }_{\left\{ {{\bf{c}}_j}, \; {\mathbf{a}}\left(l\right) \right\}} {\widetilde E}_{\rm{C}}  \\
	s.t. \; & {\bf{c}}_j \in {\mathcal{E}},   j = 1, \cdots, M! \label{DeliverSequence} \\
	& a_{k} \left(l\right) \in \left\{0, 1 \right\}, \forall l \in {\cal{L}}, k \in {\cal{K}},  \label{Binary} \\
	& R\left(l\right) \ge R_{\rm{th}}, \forall l \in {\cal{L}}, \label{Rate_Con} 
\end{align} 
\end{subequations}
where ${\bf{c}}_j$ denotes the delivery sequence,
$\mathbf{a} \left(l\right) = [ a_{1} \left(l\right), \cdots, a_{K} \left(l\right) ]^T \in {\mathbb{Z}}^{K \times 1}$ denotes the PA activation vector at $l$-th time slot, 
(\ref{DeliverSequence}) means that the UAV delivery sequence should be selected from set ${\mathcal{E}}$,
(\ref{Binary}) is the binary constraint of PA activation, 
and (\ref{Rate_Con}) requires the achievable communication rate of UAV to surpass a minimum threshold for effective transmission of controlling commands.

\section{Proposed Solutions} \label{Solution}

The optimization problem (\ref{P}) involves the UAV delivery sequence planning (an NP-hard TSP-like problem) and PA activation vector optimization (a highly coupled MINLP problem) with strong coupling, and directly solving them in a unified way faces rigorous challenges. 
Therefore, in this section, we propose a double layer optimization algorithm to minimize communication energy consumption during the UAV delivery cycle. 
Specifically, a HAO-based UAV delivery sequence planning scheme is proposed {\textit{at the outer layer}}, and a BnB-based and a ISLR-based PA activation vector optimization algorithm are proposed {\textit{at the inner layer}}. 

\subsection{Outer Layer: UAV Delivery Sequence Planning}

To reduce communication energy consumption, we aim to minimize the communication time as much as possible, which means shortening the duration $T$ of one UAV flight cycle. 
Given the UAV's constant flight speed and constant delivery speed, the way to reduce $T$ is to minimize the total flight distance within one cycle. This transformation turns the delivery sequence planning problem into a classic TSP.
The mathematical form of UAV delivery sequence planning problem can be given by 
\begin{equation}\label{P1}
	\begin{array}{l}
		({\rm{P1}})\; \mathop {\min } \limits_{\left\{ {{\bf{c}}_j} \right\}} T, \vspace{1ex}\\
		\;\; s.t. \; {\rm{(\ref{DeliverSequence})}},
	\end{array} 
\end{equation}
which is NP-hard. 
The most straightforward approach is to exhaustively search all possible routes, which amounts to $M!$ possibilities with a computational complexity of ${\cal{O}}\left( M! \right)$. Another classical method is DP, which has a computational complexity of ${\cal{O}}\left\{ M^2 2^M \right\}$ and a space complexity of ${\cal{O}}\left( M 2^M \right)$. However, when $M$ is relatively large, e.g., $M \ge 20$, the complexity of both methods becomes computationally intractable. 
While some approximation algorithms, such as the nearest neighbor greedy algorithm and GA, are capable of solving large-scale TSP instances, their solution quality tends to be unstable. 

Therefore, we propose a HAO scheme, where a GA performs coarse-grained global exploration to generate candidate solutions at the top-level, and a DP performs fine-grained local refinement of these candidates into elite solutions at the lower-level. 
The elite solutions are then fed back into the next iteration until convergence or reaching the maximum iteration threshold. 
The detailed descriptions of the HAO scheme are provided in the following subsections.  

\subsubsection{Top-level: GA-based global exploration}

In this subsection, we propose a GA-based global exploration scheme to generate candidate solutions.

\subsubsubsection{Population and chromosome initialization}

The number of population is $N$, and the number of chromosomes is $M$, which equals to the number of delivery nodes. 
In order to achieve a balance between population quality and diversity, we consider that $p_{\rm{g}}$ ($0 \le p_{\rm{g}} \le 1$) part of population is generated with nearest neighbor greedy algorithm, and $\left(1-p_{\rm{g}} \right)$ part population is generated with random algorithm. 
However, since the greedy algorithm yields a very limited number of solutions, we allocate $p_{\rm{g}} < 10\%$.

\subsubsubsection{Design of fitness function}
Since the objective is to minimize the total trajectory distance, the fitness function is defined as the reciprocal of the total distance, given by
\begin{equation}\label{Fitness}
	\begin{array}{l}
	f\left( {\bf{c}}_j \right) = \frac{1}{\left| {{{\bf{q}}_{j, 1}} - {{\bf{q}}_{\rm{S}}}} \right|+{\sum\limits_{m = 1}^{M-1} {\left| {{{\bf{q}}_{j, m}} - {{\bf{q}}_{j, m+1}}} \right| +} \left| {{{\bf{q}}_{j, M}} - {{\bf{q}}_{\rm{S}}}} \right|}},
	\end{array}
\end{equation}
where ${{\bf{c}}_j} = \left[ {{{\bf{q}}_{j,1}}, \cdots, {{\bf{q}}_{j,m}},{{\bf{q}}_{j,m + 1}}, \cdots, {{\bf{q}}_{j,M}}} \right]$ with ${{\bf{q}}_{j, m}}$ representing location of the $m$-th delivery node in the $j$-th possible delivery sequence, and $D\left( {\bf{c}}_j \right) = 1/f\left( {\bf{c}}_j \right)$. 

\subsubsubsection{Selection, crossover, and mutation}

First, consider the selection operator. The roulette wheel selection method is adopted, retaining $p_{\rm{s}}$ ($0 \le p_{\rm{s}} \le 1$) part of the individuals in each generation, while the remaining $\left( 1- p_{\rm{s}} \right)$ part of individuals are eliminated from the population.

Second, consider the crossover operator. 
The individuals retained have approximately a $p_{\rm{c}}$ ($0 \le p_{\rm{c}} \le 1$) probability of undergoing chromosomal crossover. 
The ordered crossover (OX) method is adopted, where a randomly selected segment from Parent 1 is preserved, and the remaining chromosomes are filled in the order of Parent 2.
For example, given Parent 1 and Parent 2 with the following chromosome sequences 
\begin{subequations}
	\begin{equation}
		{{\bf{c}}_1} = \left[ {{{\bf{q}}_{1,1}}, \cdots ,{{\bf{q}}_{1,m}}, \cdots , {{\bf{q}}_{1,m + I}}, \cdots ,{{\bf{q}}_{1,M}}} \right],
	\end{equation}
	\begin{equation}
		{{\bf{c}}_2} = \left[ {{{\bf{q}}_{2,1}}, \cdots ,{{\bf{q}}_{2,m}}, \cdots ,{{\bf{q}}_{2,m + I}}, \cdots ,{{\bf{q}}_{2,M}}} \right],
	\end{equation}
\end{subequations}
the resulting offspring chromosome sequence after OX is
\begin{equation}\label{offspring_OX}
{{\bf{c}}_{{\rm{ox}}}} = \left[ {\underbrace {{{\bf{q}}_{2,1'}}, \cdots }_{{\rm{from\;Parent\;2}}},\underbrace {{{\bf{q}}_{1,m}}, \cdots ,{{\bf{q}}_{1,m + I}}}_{{\rm{from\;Parent\;1}}},\underbrace { \cdots ,{{\bf{q}}_{2,M'}}}_{{\rm{from\;Parent\;2}}}} \right],
\end{equation}
where ${{\bf{q}}_{2,m'}}, m' = 1, \cdots, M$ represents the ordered sequence of remaining chromosomes in Parent 2 after removing those contained in segment $\left[{{\bf{q}}_{1,m}}, \cdots , {{\bf{q}}_{1,m + I}}\right]$.
Note that ${{\bf{q}}_{1,m}}$ is the $m$-th element in ${{\bf{c}}_{{\rm{ox}}}}$. 

Finally, consider the mutation operator. 
The probability of mutation is $p_{\rm{m}}$ ($0 \le p_{\rm{m}} \le 1$).
The inversion mutation (IM) method is adopted, where two positions are randomly selected and the chromosomal sequence between them is reversed to enhance local search capability. 
For example, given a parent individual with the following chromosome sequence
\begin{equation}
	{{\bf{c}}_j} = \left[ {{{\bf{q}}_{j,1}}, \cdots , \underbrace {{{\bf{q}}_{j,m}}, \cdots , {{\bf{q}}_{j,m + I}}}, \cdots ,{{\bf{q}}_{j,M}}} \right],
\end{equation}
the resulting offspring chromosome sequence after IM is
\begin{equation} \label{offspring_IM}
	{{\bf{c}}_{\rm{IM}}} = \left[ {{{\bf{q}}_{j,1}}, \cdots ,\underbrace {{{\bf{q}}_{j,m + I}}, {{\bf{q}}_{j,m + I-1}},\cdots , {{\bf{q}}_{j,m}}}, \cdots ,{{\bf{q}}_{j,M}}} \right].
\end{equation}

\begin{algorithm}[t]
	\caption{GA-based global exploration}
	\label{alg:GA}
	\begin{algorithmic}[1]
		\begin{small}
			\REQUIRE 
			Initial number of individuals in the population, $N$; 
			Number of chromosomes, $M$ (equals to the number of delivery nodes); 
			Probability of selection, crossover and mutation, $p_{\rm{s}}$, $p_{\rm{c}}$ and $p_{\rm{m}}$, respectively; 
			Number of evolutionary generations, $G_{\max}$; 
			Iteration index of HAO scheme, $I$; 
			\ENSURE
			Candidate solutions, ${\bf{c}}_j, j = 1, 2, \cdots, N/10$.
			\STATE {\textit{Initialization:}} $G=1$; 
			\IF{$I=1$} 
			\STATE Generate $p_{\rm{g}}N$ individuals via greedy algorithm, and $\left(1-p_{\rm{g}}\right)N$ individuals via random generation; \\
			\ELSE
			\STATE Retain elite solutions obtained by DP, generate $9/10 \times p_{\rm{g}}N$ individuals via greedy algorithm, and $9/10 \times \left(1-p_{\rm{g}}\right)N$ individuals via random generation;
			\ENDIF
			\WHILE{$G \le G_{\max}$}
			\STATE {\textit{Fitness calculation:}} Calculate fitness according to (\ref{Fitness}) for each individual in the population, and obtain $f \left( {\bf{c}}_j \right), j = 1, \cdots, N$; \\
			\STATE {\textit{Selection:}} Retain $p_{\rm{s}} N$ individuals using roulette wheel selection and discard $\left(1 - p_{\rm{s}}\right) N$ individuals; \\
			\STATE {\textit{Crossover:}} Randomly select $p_{\rm{c}} p_{\rm{s}} N$ individuals for chromosome crossover as specified in  (\ref{offspring_OX}), where Parent 1 is selected from elite candidates;  \\
			\STATE {\textit{Mutation:}} Randomly select $p_{\rm{m}} p_{\rm{s}} N$ individuals for mutation as specified in  (\ref{offspring_IM}); \\
			\STATE {\textit{Regeneration population:}} Recalculate the fitness of updated individuals. According to the sorted order of the updated fitness values, the top 1/2 of the population (ranked by fitness) are retained, and the remaining 1/2 of the population are generated randomly, ensuring that the total population size remains $N$.\\
			\STATE $G \gets G+1$;
			\ENDWHILE		
			\STATE {\textit{Candidate solutions:}} 1/10 of the individuals with the highest fitness are selected as candidate solutions.
		\end{small}
	\end{algorithmic}		
\end{algorithm}

\subsubsubsection{Generation of candidate solution}

Through $G_{\max}$ generations of evolution, the optimal individual from each generation along with select diverse individuals are retained to form a candidate solution set ${\cal{C}} \triangleq \left\{ {\bf{c}}_1, \cdots, {\bf{c}}_{N'} \right\}$. 
The size of this solution set is maintained below the population size, i.e., ${N'} \le N$ to ensure both solution diversity and quality. 
The details of GA-based global exploration are summarized in {{Algorithm {\ref{alg:GA}}}}.

\subsubsection{Lower-level: DP-based Local Refinement} 

Through the GA-based global exploration, a set of candidate solutions is obtained, where each candidate solution represents a potential UAV delivery sequence. 
Building upon this, we employ DP method to perform local refinement on each candidate path, enhancing solution quality through exact subpath computations. 
Therefore, we first divide the initial path ${{\bf{c}}_j} \in {\cal{C}}$ into $\left( b +1\right)$ distinct subpaths of length $a$, where $b = \left \lfloor \frac{M}{a} \right \rfloor$. Each subpath contains $\left( a +1\right)$ delivery nodes, denoted by ${\rm{subpath}}\underline{\;}i, i = 0,1, \cdots, b$, given by
\begin{subequations}\label{Subpath}
	\begin{align}
	& {\rm{subpath}}\underline{\;}0 =  \left[{{\bf{q}}_{\rm{S}}},{{\bf{q}}_1},{{\bf{q}}_2}, \cdots ,{{\bf{q}}_a}\right], \\
	& {\rm{subpath}}\underline{\;}1 =  \left[{{\bf{q}}_a},{{\bf{q}}_{a+1}}, \cdots ,{{\bf{q}}_{2a}}\right], \\
	& \cdots \nonumber \\
	& {\rm{subpath}}\underline{\;}b =  \left[{{\bf{q}}_{ba}},{{\bf{q}}_{ba+1}}, \cdots ,{{\bf{q}}_{M}} ,{{\bf{q}}_{S}} \right],
	\end{align}
\end{subequations}
where the ending node of ${\rm{subpath}}\underline{\;}i$ is the starting node of  ${\rm{subpath}}\underline{\;}{\left(i+1\right)}$.
This is  to avoid long distances at the junctions when multiple subpaths are joined to form the entire path.

The set of visited delivery nodes is denoted by ${\cal{S}}$, and the state $\left({\cal{S}}, {\bf{q}}_u \right)$ indicates that the set of visited nodes is ${\cal{S}}$ and the current location is at node ${\bf{q}}_u$.
Define $dp\left[ { {\cal{S}} } \right] \left[ {\bf{q}}_u \right]$ as the minimum cost to visit all delivery nodes in subset ${\cal{S}}$, ending at node ${\bf{q}}_u$. 
For the state $\left({\cal{S}}, {\bf{q}}_u \right)$, when transitioning to the next node ${\bf{q}}_v $, we have
\begin{equation}
	\begin{aligned}
	dp\left[ {\cal{S}}' \right] \left[ {\bf{q}}_v \right] = & dp\left[ { {\cal{S}} \cup \left\{ {\bf{q}}_v \right\}} \right] \left[ {\bf{q}}_v \right]  \\ = &  \min \Big( dp\left[ {\cal{S}} \right]\left[ {\bf{q}}_u \right] + \left| {\bf{q}}_u - {\bf{q}}_v \right| \Big), 
	\end{aligned}
\end{equation}
where ${\cal{S}}' \triangleq {\cal{S}} \cup \left\{ {\bf{q}}_v \right\}$ represents the updated set of visited delivery nodes.

\begin{algorithm}[t]
	\caption{DP-based local refinement}
	\label{alg:DP}
	\begin{algorithmic}[1]
		\begin{small}
		\REQUIRE 
		Candidate solutions obtained by GA-based exploration, ${\bf{c}}_j, j = 1, 2, \cdots$; 		
		\ENSURE
		Refined candidate solutions, ${\bf{c}}_{j}^{\rm{ref}}, j = 1, 2, \cdots$.
		\FOR{$\forall {\bf{c}}_j, j = 1, 2, \cdots$}
		\STATE {\textit{Candidate decomposition:}} Decompose ${\bf{c}}_j$ into multiple subpaths according to the method described in (\ref{Subpath});\\
		\FOR{${\rm{subpath}}\underline{\;}i, i = 0,1, \cdots, b$}
		\STATE {\textit{Subpath optimization:}} Dynamic programming for subpath optimization according to S\ref{S1}-S\ref{S7};\\
		\ENDFOR
		\STATE {\textit{Recombination:}} Combine the optimized subpaths in their original sequence to form the refined  solution ${\bf{c}}_j^{\rm{ref}} =$ $\big[ {\rm{subpath}} \underline{\;}0', {\rm{subpath}} \underline{\;}1', \cdots, {\rm{subpath}} \underline{\;}b'\big]$. \\
		\IF{$D\left({\bf{c}}_j^{\rm{ref}}\right) > D\left({\bf{c}}_j\right)$}
		\STATE{${\bf{c}}_j^{\rm{ref}} = {\bf{c}}_j$;} \\
		\ENDIF
		\ENDFOR
		\end{small}
	\end{algorithmic}		
\end{algorithm}

Based on the aforementioned definition of state and state transition rules, we present the DP procedure as follows
\begin{enumerate}[S1.]
	\item \label{S1} For each ${\rm{subpath}} \underline{\;}i$, denote the starting node by ${\bf{q}}_{i, 0}$, denote the ending point by ${\bf{q}}_{i, a}$, and initialize the set of visited nodes ${\cal{S}} = \left\{{\bf{q}}_{i, 0}\right\}$;
	\item \label{S2} For each ${\rm{subpath}} \underline{\;}i$, initialize the path length by setting $dp\left[ \left\{ {\bf{q}}_{i, 0} \right\}\right] \left[ {\bf{q}}_{i, 0} \right] =0$, and setting all other path lengths $dp\left[ \left\{ {\bf{q}}_{i, 0} \right\}\right] \left[ {\bf{q}}_{i, j} \right] =$ $+ \infty$, $j \ne 0$;
	\item \label{S3} Starting from state $\left( {\cal{S}}, {\bf{q}}_{i, 0} \right)$, traverse all possible next nodes ${\bf{q}}_{i, u} \notin {\cal{S}} \cup \left\{{\bf{q}}_{i, a}\right\}$, update the set of visited nodes ${\cal{S}} \gets {\cal{S}} \cup \left\{ {\bf{q}}_{i, u}\right\}$, update the state $\left( {\cal{S}}, {\bf{q}}_{i, u} \right)$ and calculate the path length $dp\left[ {\cal{S}} \right] \left[ {\bf{q}}_{i, u} \right]$ for each newly added node;  
	\item \label{S4} Starting from each state $\left( {\cal{S}}, {\bf{q}}_{i, u} \right)$, traverse all possible next nodes ${\bf{q}}_{i, v} \notin {\cal{S}} \cup \left\{{\bf{q}}_{i, a}\right\}$, update the set of visited nodes ${\cal{S}} \gets {\cal{S}} \cup \left\{ {\bf{q}}_{i, v}\right\}$, update the state $\left( {\cal{S}}, {\bf{q}}_{i, v} \right)$ and calculate the path length $dp\left[{\cal{S}} \right] \left[ {\bf{q}}_{i, v} \right]$ for each newly added node;
	\item \label{S5} Repeat the S{\ref{S4}} until set ${\cal{S}}$ includes all delivery nodes of ${\rm{subpath}} \underline{\;}i$ (except the ending node ${\bf{q}}_{i, a}$);
	\item \label{S6} Starting from all the possible current state obtained by S{\ref{S5}}, visit the ending node ${\bf{q}}_{i, a}$, update the set of visited nodes ${\cal{S}} \gets {\cal{S}} \cup \left\{ {\bf{q}}_{i, a}\right\}$, update the state $\left( {\cal{S}}, {\bf{q}}_{i, a} \right)$, and calculate the path length $dp\left[{\cal{S}} \right] \left[ {\bf{q}}_{i, a} \right]$; 
	\item \label{S7} Find the shortest path from the starting node to the ending node based on the DP results, and obtained the optimized ${\rm{subpath}} \underline{\;}i'$. 
\end{enumerate}

The optimized subpaths $\big[ {\rm{subpath}} \underline{\;}0',$ $ {\rm{subpath}} \underline{\;}1', \cdots, $ ${\rm{subpath}} \underline{\;}b'\big]$ are sequentially recombined to form a refined entire-path solution ${\bf{c}}_{j}^{\rm{ref}}$. 
If the distance of the refined path is shorter than that of the original path, i.e., $D\left( {\bf{c}}_{j} ^{\rm{ref}} \right) < D\left( {\bf{c}}_{j} \right)$, the original path is replaced with the new one, otherwise keep the original path.
The details of DP-based local refinement are summarized in {{Algorithm {\ref{alg:DP}}}}.

\subsubsection{HAO-based UAV Delivery Sequence Planning Scheme}

By combining GA-based global exploration at the top-level and DP-based local refinement at the lower-level, this approach effectively balances search breadth and solution accuracy. 
The process iteratively alternates between these two phases until achieving convergence or the maximum iteration times, ultimately obtaining high-quality solution.
The details of HAO-based UAV delivery sequence planning scheme are summarized in {{Algorithm {\ref{alg:HAO}}}}. 

\begin{algorithm}[t]
	\caption{HAO-based UAV delivery sequence planning}
	\label{alg:HAO}
	\begin{algorithmic}[1]
		\begin{small}
		\REQUIRE 
		Maximum number of iteration, $I_{\max}$. \\
		\ENSURE
		Optimal delivery sequence, ${\bf{c}}_{\rm{opt}}$.
		\STATE {\bf{Initialization:}} $I = 1$, $D \left( {\bf{c}}_{\rm{opt}} \right)= +\infty$. 
		\WHILE{$I \le I_{\max}$}
		\STATE {\textit{Top-level:}} Carry out GA-based global exploration according to Algorithm {\ref{alg:GA}}, obtain candidate solutions ${\bf{c}}_j, j = 1, 2, \cdots$;
		\STATE {\textit{Lower-level:}} Carry out DP-based refinement according to Algorithm {\ref{alg:DP}} to further optimize candidates from GA, obtain refined candidate solutions ${\bf{c}}_{j}^{\rm{ref}}, j = 1, 2, \cdots$;
		\STATE Calculate the path length of each refined candidate solution $D \left( {\bf{c}}_{j} ^{\rm{ref}} \right)$;
		\STATE Obtain the optimal solution ${\bf{c}}_{\rm{opt}}^{I}$ with the shortest path length, i.e., $D\left( {\bf{c}}_{\rm{opt}}^{I} \right) = \mathop {\min }\limits_{j = 1,2, \cdots } D\left( {{\bf{c}}_j^{{\rm{ref}}}} \right) $;
		\IF {$D \left( {\bf{c}}_{\rm{opt}}^{I} \right) < D \left( {\bf{c}}_{\rm{opt}} \right)$}
		\STATE ${\bf{c}}_{\rm{opt}} = {\bf{c}}_{\rm{opt}}^{I}$, $D \left( {\bf{c}} _{\rm{opt}} \right) = D \left( {\bf{c}} _{\rm{opt}}^{I} \right)$;
		\ENDIF
		\STATE $I \gets I+1$;
		\STATE Add the new candidates ${\bf{c}}_{j}^{\rm{ref}}, j = 1, 2, \cdots$ obtained through DP to the existing population, and obtain the updated population;
		\ENDWHILE
		\end{small}
	\end{algorithmic}
\end{algorithm}

\subsection{Inner Layer (Optimal Solution): BnB-based PA Activation Vector Optimization}

With given location of UAV ${\bm{\psi}} _{{\rm{U}}} \left(l\right)$ at time slot $l$, we aim to minimize the energy consumption of communication through PA activation vector optimization while satisfying the minimum rate constraint. 
Consider an ideal scenario where the activation and deactivation of antennas can be completed instantaneously. 
By minimizing the energy consumption of each time slot, the energy consumption of the entire flight cycle is minimized, 
thus the optimization problem can be formulated as
\begin{equation}\label{P2}
\begin{array}{l}
	({\rm{P2}})\; \mathop {\min } \limits_{\left\{ {\mathbf{a}}\left(l\right) \right\}} {\widetilde E}_{\rm{C}} \left(l\right), \vspace{1ex} \\
	\;\; s.t. \; {\rm{(\ref{Binary}), (\ref{Rate_Con})}}. 
\end{array} 
\end{equation}

The minimum energy consumption of communication in (\ref{P2}), i.e., ${\widetilde E}^*_{\rm{C}} \left(l\right)$, is always achieved at the lower bound given by constraint (\ref{Rate_Con}), yielding
\begin{equation}
	{\widetilde E}^*_{\rm{C}} \left(l\right) = { P_{\rm{C}}^* \left(l\right) \tau },
\end{equation}
where 
\begin{equation}\label{P_C}
P^*_{\rm{C}} \left(l\right) = \frac{\left(2^{R_{\rm{th}}} - 1\right){\sigma^2}} { \left| {\bf{h}}^H \left( \mathbf{x}, l \right) {\bf{G}}\left( {\bm{\beta}} \left(l\right), \mathbf{x} \right) {\mathbf{a}} \left(l\right) \right|^2 }.
\end{equation}
Thus, problem (\ref{P2}) is equivalent to maximizing $\frac{1}{P^*_{\rm{C}} \left(l \right)}$ by solving the following problem
\begin{equation}\label{P2_1}
\begin{array}{l}
	 \mathop {\max } \limits_{\left\{ {\mathbf{a}}\left(l\right) \right\}} f \left({\mathbf{a}}\left(l\right) \right) =  \rho \left| \left[ {\bf{h}} \left( \mathbf{x}, l \right)\right]^H  {\bf{G}}\left( {\bm{\beta}} \left(l\right), \mathbf{x} \right) {\mathbf{a}} \left(l\right) \right|^2, \vspace{1ex}\\
	\;\; s.t. \; {\rm{(\ref{Binary})}}, 
\end{array} 
\end{equation}
where $\rho = \frac {1}{\left(2^{R_{\rm{th}}} - 1\right){\sigma^2}}$.

We observe that the optimization variable ${\mathbf{a}}\left(l\right)$ in (\ref{P2_1}) has a finite set of possible values. Each element $a_k \left(l\right)$ in the vector a can only take the value 0 or 1. Therefore, by exhaustively enumerating all $2^K$ possible combinations, we can obtain the optimal solution ${\mathbf{a}}^*\left(l\right)$ to (\ref{P2_1}). 
However, when the number of pre-mounted PAs, i.e., $K$, is large, the complexity of exhaustive search is unacceptable, which motivates us to explore an efficient approach to obtain the globally optimal solution of (\ref{P2_1}). 
Thus, we aim to employ the BnB method to obtain the globally optimal solution.

By splitting the full solution space ${\bf x} \in {\cal{B}}_{\rm{ALL}}$ into multiple subspaces denoted as ${\cal{S}} = \left\{{\cal{B}}_1, \cdots, {\cal{B}}_S \right\}$ ({\textit{branching}}) and computing the objective function's boundaries for each individual subspace ({\textit{bounding}}), the BnB approach can efficiently rule out subspaces that have no possibility of containing the optimal solution, thus avoiding a full exhaustive search. 
These subspaces ${\cal{B}}_s \in {\cal{S}}$ are commonly called ``{\textit{boxes}}". For every box, the BnB method addresses a convex relaxation problem. This step serves two key purposes: verifying whether the subspace is feasible and determining the lower and upper bounds for the original nonconvex problem. As the process of dividing subspaces continues and the size of each box shrinks, the global upper and lower bounds become increasingly precise. Over time, these bounds converge toward the global optimal solution \cite{lawler1966branch, balakrishnan1991branch}.

To solve the $K$-dimensional PA activation vector optimization problem, we define a box ${\cal {B}}$, which is a $K$-dimensional hyperrectangle of variable ${\bf{x}}$ with lower bound $\underline {\bf{b}}  = {\left[ {{{\underline b }_1},{{\underline b }_2}, \cdots ,{{\underline b }_K}} \right]^T}$ and upper bound $\overline {\bf{b}}  = {\left[ {{{\overline b }_1},{{\overline b }_2}, \cdots ,{{\overline b }_K}} \right]^T}$, which is defined as
\begin{equation}
	{\cal {B}} \triangleq \left[ \underline {\bf{b}}, \overline {\bf{b}} \right] = \left\{ {{\bf{b}} \in {{\mathbb{R}}^{K \times 1}}\left| {{{\underline b }_i} \le {x_i} \le {{\overline b }_{i,}} \; \forall i = 1, \cdots ,K} \right.} \right\}. 
\end{equation}
BnB exploits bounding estimation functions $f_{\rm{LB}}\left({\cal {B}}\right)$ and $f_{\rm{UB}}\left({\cal {B}}\right)$ to evaluate the lower and upper bounds of the local optimal objective function value $f^*\left({\cal {B}}\right)$ within each box ${\cal {B}}$, such that 
\begin{enumerate}[i)]
	\item $f_{\rm{LB}}\left({\cal {B}}\right) \le f^*\left({\cal {B}}\right) \le f_{\rm{UB}}\left({\cal {B}}\right)$;
	\item $f_{\rm{UB}}\left({\cal {B}}\right) - f_{\rm{LB}}\left({\cal {B}}\right) \to 0$ as B shrinks to a point.
\end{enumerate}

As more stringent bounds are derived from smaller feasible subspaces, the global upper bound undergoes a gradual reduction. 
Furthermore, the global lower bound is refined repeatedly through two key steps: first, pruning redundant boxes that hold no potential for containing the optimal solution, and second, shrinking the boxes that remain.
When the sizes of the boxes within the set \({\cal{S}}\) decrease over $I$ iterations, the bound gap $f_{\rm{UB}}\left({\cal {B}}\right) - f_{\rm{LB}}\left({\cal {B}}\right)$ converges to 0. At this point, the globally optimal value $f^*\left({\cal {B}}\right)$ can be approximated as
\begin{equation}
\begin{array}{l}
{\rm{GLB}}\left[1\right] \le \cdots \le {\rm{GLB}}\left[I\right] \le \vspace{1ex}\\
 \;\;\;\; f\left({\bf{x}}^*\right) \le {\rm{GUB}}\left[I\right] \le \cdots \le {\rm{GUB}}\left[1\right],
\end{array}
\end{equation}
where ${\bf{x}}^*$ is the optimal solution of the original problem. 

To effectively estimate the bound, we first build a convex relaxation of problem (\ref{P2_1}). Subsequently, we design a BnB algorithm to acquire the optimal solution. 

\subsubsection{Convex Relaxation}

To begin with, we relax the binary constraints in (\ref{P2_1}) to continuous constraints. Thus, the mathematical formulation of the optimization problem is
\begin{subequations}\label{P_FixedPAnum}
	\begin{align}
		& \mathop {\max }_{\left\{ {\mathbf{a}}\left(l\right) \right\}} f \left({\mathbf{a}}\left(l\right) \right) = \rho {\left| {{\bf{h}}} ^H \left( \mathbf{x}, l \right) {\bf{G}}\left( {\bm{\beta}} \left(l\right), \mathbf{x} \right) {\mathbf{a}} \left(l\right) \right|^2 }, \\
		& s.t. \; a_{k} \left(l\right) \in \left[0, 1\right], \forall l \in {\cal{L}}, k \in {\cal{K}}. \label{Binary_relax} 
	\end{align}
\end{subequations}

From Eq. ({\ref{PowerModel}}), it can be seen that the power radiation ratio vector ${\bm{\beta}} \left(l\right)$ is independent of the specific locations of the PAs, and is determined by the PA activation vector. 
Substituting ${\bf{G}}\left( {\bm{\beta}} \left(l\right), \mathbf{x} \right)$ with ${\rm{diag}} \left({\bm{\beta}} \left(l\right)\right) {\widetilde{\bf{G}}}\left( \mathbf{x}, l \right) $, the objective function of (\ref{P_FixedPAnum}) can be rewritten as
\begin{equation}
\begin{array}{l}
	f \left({\mathbf{a}}\left(l\right) \right)  = \rho {\left| {{\bf{h}}} ^H \left( \mathbf{x}, l \right) {\rm{diag}} \left({\bm{\beta}} \left(l\right)\right) {\widetilde{\bf{G}}}\left( \mathbf{x}, l \right) {\mathbf{a}} \left(l\right) \right|^2 }, \vspace{1ex}\\
	\Rightarrow f \left({\mathbf{a}}\right)  = \rho {{\bf{h}}} ^H {\rm{diag}} \left({\bm{\beta}} \right) {\widetilde{\bf{G}}} {\mathbf{a}} {\mathbf{a}}^T {\widetilde{\bf{G}}} ^H  {\rm{diag}} \left({\bm{\beta}} \right)  {{\bf{h}}},
\end{array}
\end{equation}
where $\widetilde{\bf{G}} = {\rm{diag}}\left( \left[ {\tilde{g}}_1, \cdots, {\tilde{g}}_k, \cdots, {\tilde{g}}_K \right]^T \right) \in {\mathbb{C}}^{K \times K}$, and the indices $\mathbf{x}$ and $l$ are omitted to simplify the notation. 
We newly introduce a variable ${\mathbf{A}} \in {\mathbb{R}}^{K \times K}$, which satisfies
\begin{equation}\label{Aaa}
	{\mathbf{A}} = {\mathbf{a}} {\mathbf{a}}^T = \left[a_k a_{k'}\right]. 
\end{equation}
The convex hull corresponding to the bilinear term $\left[a_k a_{k'}\right]$ can be derived via the McCormick envelope. On this basis, the convex relaxation form of the constraint in (\ref{Aaa}) can be expressed as
\cite{mccormick1976computability}
\begin{subequations}\label{McCormick_A}
	\begin{align}
		& {\bf{A}} \ge {\underline {\bf{a}}} \; {\bf{a}}^T + {\bf{a}} \; {\underline {\bf{a}}}^T  - {\underline {\bf{a}}} \; {\underline {\bf{a}}}^T, \\
		& {\bf{A}} \ge {\overline {\bf{a}}} \; {\bf{a}}^T + {\bf{a}} \; {\overline {\bf{a}}}^T  - {\overline {\bf{a}}} \; {\overline {\bf{a}}}^T, \\
		& {\bf{A}} \le {\overline {\bf{a}}} \; {\bf{a}}^T + {\bf{a}} \; {\underline {\bf{a}}}^T  - {\overline {\bf{a}}} \; {\underline {\bf{a}}}^T, \\
		& {\bf{A}} \le {\underline {\bf{a}}} \; {\bf{a}}^T + {\bf{a}}\; {\overline {\bf{a}}}^T - {\underline {\bf{a}}} \; {\overline {\bf{a}}}^T,
	\end{align}
\end{subequations}
where 
\begin{equation}\label{Bounds_a}
{\underline {\bf{a}}} \le {{\bf{a}}} \le {\overline {\bf{a}}}.
\end{equation}

\begin{myRemark}
	{\textit{When ${\underline {\bf{a}}} = {\overline {\bf{a}}}$, the equalities in constraints (\ref{McCormick_A}) are satisfied. Consequently, the McCormick envelope reduces to the bilinear constraints.}}
\end{myRemark}

In addition, the complex coupling relationship between the power radiation ratio vector $\bm{\beta}$ and the PA activation vector $\bf{a}$ renders the objective function nonconvex and difficult to handle. To establish an upper bound for the objective function, we relax all elements in the power radiation ratio vector $\bm{\beta}$ to its first element, $\beta_1$.
Hence, the nonconvex (\ref{P2_1}) can be relaxed into the following convex optimization problem:
\begin{equation}\label{RelaxedP2_1}
\begin{array}{l}
	\mathop {\max } \limits_{\left\{{\mathbf{A}}, {\mathbf{a}} \right\}} \rho \beta_1^2 {{\bf{h}}} ^H {\widetilde{\bf{G}}} {\mathbf{A}} {\widetilde{\bf{G}}} ^H  {{\bf{h}}}, \vspace{1ex}\\
	\;\; s.t. \; {\rm{(\ref{Binary_relax}), (\ref{McCormick_A}), (\ref{Bounds_a})}}.
\end{array} 
\end{equation}
Solve the relaxed problem (\ref{RelaxedP2_1}) to obtain the optimal solution ${\bf{x}}_{\rm{relax}} = {\left\{{\mathbf{A}}_{\rm{relax}}, {\mathbf{a}}_{\rm{relax}} \right\}}$ and optimal objective function value $f_{\text{relax}}^*$. Since this optimal value is obtained under relaxed conditions that loosen the objective function and constraints of the original problem, its result must be no smaller than the optimal solution of the original problem. 
Therefore, $f_{\text{relax}}^*$ can serve as an upper bound for the original problem, which will be used for pruning operations in the subsequent execution. 

\subsubsection{BnB Algorithm}

To find the optimal solution for problem (\ref{P2_1}), we carry out branching operations on the discrete variables ${\bf{a}} = \left[a_k \right]$.
As a result, the $K$-dimensional variables used for branching are defined as ${\bf{b}} ={\bf{a}} \in {\mathbb R}^{K \times 1}$. Based on the given definitions, the initial solution space region ${\cal{B}}_{\rm{ALL}}$ can be determined by setting the lower bound as $\underline {\bf{b}} = {\bf{0}}_{K \times 1}$ and the upper bound as $\overline {\bf{b}} = {\bf{1}}_{K \times 1}$. 
In each iteration of the BnB process, three core steps, i.e., branching, bounding, and pruning, are executed sequentially.

\subsubsubsection{Branching}

At each iteration, a box ${\cal{B}}_o$ is selected from the candidate set ${\cal{S}}$, and this box is split into two child boxes ${\cal{B}}_-$ and ${\cal{B}}_+$ along a specific edge $e$, where $e \in \left\{1, \cdots, K \right\}$. 
The box selection for branching follows the best-bound-first rule referenced in {\cite{balakrishnan1991branch}}. Specifically, the box that attains the optimal upper bound is chosen for the branching process
\begin{equation}\label{SelectBox}
	{{\cal B}_o} = \mathop {\arg \max }\limits_{{\cal B} \in {\cal S}} {f_{{\rm{UB}}}}\left( {\cal B} \right).
\end{equation}
In addition, the edge selection process follows the maximum-length-first rule. For the selected box ${\cal{B}}_o$, it is split evenly into two sub-boxes ${\cal{B}}_-$ and ${\cal{B}}_+$ along the longest edge
\begin{equation}\label{SelectEdge}
	e = \mathop {\arg \max }\limits_{i \in \left\{ {1,2, \cdots ,K} \right\}} \left| {{{\overline b }_i} - {{\underline b }_i}} \right|.
\end{equation}
Generate two branches by setting $ b_e = 0$ and $ b_e = 1$ respectively, and the remaining elements remain unchanged. 
The resultant boxes can be defined as ${\cal{B}}_- = \left[{\underline{\bf{b}}}_{-}, {\overline{\bf{b}}}_{-}\right]$ and ${\cal{B}}_+ = \left[{\underline{\bf{b}}}_{+}, {\overline{\bf{b}}}_{+}\right]$, where  ${\underline{\bf{b}}}_{-}$, ${\overline{\bf{b}}}_{-}$, ${\underline{\bf{b}}}_{+}$ and ${\overline{\bf{b}}}_{+}$ are given by 
\begin{subequations}\label{ChildrenBox}
\begin{align}
& {\underline{\bf{b}}}_{-} = \left[{\underline{b}}_1, {\underline{b}}_2, \cdots, {\underline{b}}_{e-1}, 0, {\underline{b}}_{e+1}, \cdots, {\underline{b}}_K\right]^T,\\
& {\overline{\bf{b}}}_{-} = \left[{\overline{b}}_1, {\overline{b}}_2, \cdots, {\overline{b}}_{e-1}, 0, {\overline{b}}_{e+1}, \cdots, {\overline{b}}_K\right]^T,\\
& {\underline{\bf{b}}}_{+} = \left[{\underline{b}}_1, {\underline{b}}_2, \cdots, {\underline{b}}_{e-1}, 1, {\underline{b}}_{e+1}, \cdots, {\underline{b}}_K\right]^T,\\
& {\overline{\bf{b}}}_{+} = \left[{\overline{b}}_1, {\overline{b}}_2, \cdots, {\overline{b}}_{e-1}, 1, {\overline{b}}_{e+1}, \cdots, {\overline{b}}_K\right]^T.
\end{align}
\end{subequations}
Then, we can update the candidate box list $\cal S$ with
\begin{equation}\label{UpdateS}
	{\cal S} \leftarrow {\cal S} \setminus \left\{ {\cal{B}}_o\right\} \cup \left\{ {\cal{B}}_-, {\cal{B}}_+\right\}. 
\end{equation}
Through this branching approach, the solution space of the original problem is continuously subdivided to form a search tree, where each node represents a partial solution to a subproblem.

\subsubsubsection{Bounding}

We assess the bounds of $f^*$ for each child box ${\cal{B}} \in \left\{ {\cal{B}}_-, {\cal{B}}_+\right\}$. Let ${\bf{x}}_{\rm{relax}} = {\left\{{\mathbf{A}}_{\rm{relax}}, {\mathbf{a}}_{\rm{relax}} \right\}}$ denote the optimal solution of the relaxed problem (\ref{RelaxedP2_1}) within the box ${\cal{B}}$. 
The optimal value derived from the relaxed problem (\ref{RelaxedP2_1}) serves as an upper bound for the original problem (\ref{P2_1}), specifically
\begin{equation}\label{f_UB}
{f_{{\rm{UB}}}}\left( {\cal B} \right) = f\left( {\bf{a}}_{\rm{relax}} \right) = \rho {\left| {{\bf{h}}} ^H  {\widetilde{\bf{G}}} {\mathbf{a}}_{\rm{relax}} \right|^2 }. 
\end{equation}
The global upper bound ${\rm{GUB}}$ is then updated by the maximum ${f_{{\rm{UB}}}}\left( {\cal B}' \right)$ over all candidate boxes
\begin{equation}\label{GUB}
{\rm{GUB}} = \mathop {\max }\limits_{{\cal B}' \in {\cal S}} {f_{{\rm{UB}}}}\left( {\cal B}' \right). 
\end{equation}

By projecting the relaxed PA activation solution ${\bf{a}}_{\rm{relax}}$ into binary variables ${\bf{a}}_{\rm{prj}}$ via rounding operations, we can further derive a feasible solution for the original problem. On this basis, the lower bound can be evaluated as
\begin{equation}\label{f_LB}
	{f_{{\rm{LB}}}}\left( {\cal B} \right) = f \left( {\bf{a}}_{\rm{prj}} \right) = \rho {\left| {{\bf{h}}} ^H  {\widetilde{\bf{G}}} {\mathbf{a}}_{\rm{prj}} \right|^2 }. 
\end{equation}

The global lower bound ${\rm{GLB}}$ can be updated by the best feasible solution currently found, given by
\begin{equation}\label{GLB}
{\rm{GLB}} \leftarrow \max \left\{ {{\rm{GLB}},{f_{{\rm{LB}}}}\left( {{{\cal B}_ - }} \right),{f_{{\rm{LB}}}}\left( {{{\cal B}_ + }} \right)} \right\}.
\end{equation}

\subsubsubsection{Pruning} 
Boxes that cannot contain optimal solution can be identified and pruned, which accelerates the convergence process without compromising global optimality. Specifically, the following types of boxes can be pruned from $\cal S$
\begin{itemize}
	\item {\textit{Infeasible boxes:}} If the relaxed problem (\ref{RelaxedP2_1}) has no feasible solution within the range of ${\cal B}$, then ${\cal B}$ is also infeasible for the original problem (\ref{P2_1}) and can be pruned.
	\item {\textit{Fathomed boxes:}} A box ${\cal B}$ can be pruned when it is fathomed (i.e., fully explored). This occurs in two scenarios, either the local upper and lower bounds meet the condition
	\begin{subequations}\label{Fathomed}
		\begin{equation}
			{f_{{\rm{UB}}}}\left( {\cal B} \right) - {f_{{\rm{LB}}}}\left( {\cal B} \right) \le \varepsilon, \; \forall {\cal B} \in \left\{ {{{\cal B}_ - },{{\cal B}_ + }} \right\},
		\end{equation}
		or the space region of ${\cal B}$ satisfies
		\begin{equation}
			{\overline{\bf{b}}} - {\underline{\bf{b}}} \le \varepsilon {\bf{1}}, \; \forall {\cal B} \in \left\{ {{{\cal B}_ - },{{\cal B}_ + }} \right\},
		\end{equation}
	\end{subequations}
	where $\varepsilon \ge 0$ is a very small constant.
	\item {\textit{Nonoptimal boxes:}} For any box ${\cal B}'$ in ${\cal S}$, if its local upper bound $f_{\rm{UB}} \left({\cal B}'\right)$ is smaller than the global lower bound ${\rm{GLB}}$, i.e.,
	\begin{equation}\label{Nonoptimal}
		{f_{\rm{UB}}}\left( {{\cal B}'} \right) < {\rm{GLB}}, \; \forall {\cal B}' \in {\cal S},
	\end{equation}
	${\cal B}'$ definitely does not contain the optimal solution and can be pruned.
\end{itemize}

\begin{algorithm}[!t]
\caption{BnB-based PA activation optimization algorithm}
\label{alg:BnB}
\begin{algorithmic}[1]
	\begin{small}
		\REQUIRE tolerance threshold ${\epsilon}$;
		\ENSURE optimal PA activation vector ${\bf{a}}^*$, optimal objective function value $f^*$;
		\STATE {\bf{Initialization:}}
		\STATE Set PA activation vector ${\bf{a}} = {\bf{0}}_{K \times 1}$;
		\STATE Set box ${\cal {B}} = \left[ \underline {\bf{b}}, \overline {\bf{b}} \right] = \left[ {\bf{0}}_{K \times 1}, {\bf{1}}_{K \times 1} \right]$, box list ${\cal {S}} = \left\{ {\cal {B}} \right\}$;
		\STATE Set ${\rm{GUB}} = 0$, ${\rm{GLB}} = - \infty$, $f^* = - \infty$;
		\WHILE{${\cal S} \ne \emptyset$ and ${\rm{GUB}} - {\rm{GLB}} > \epsilon$}
		\STATE {\textit{Branching:}}
		\STATE Select ranching box according to (\ref{SelectBox}), and select the branching edge according to (\ref{SelectEdge});
		\STATE Obtain children boxes ${\cal B}_-$ and $ {\cal B}_+$ according to (\ref{ChildrenBox});
		\STATE Update the box list according to (\ref{UpdateS});
		\FOR{${\cal B} = \left\{{\cal B}_-, {\cal B}_+\right\}$}
		\STATE {\textit{Bounding:}}
		\STATE Solve the relaxed optimization problem (\ref{RelaxedP2_1});
		\IF{(\ref{RelaxedP2_1}) is infeasible within box ${\cal B}$}
		\STATE Prune branch ${\cal B}$;
		\ELSE
		\STATE Obtain ${\bf{x}}_{\rm{relax}} = {\left\{{\mathbf{A}}_{\rm{relax}}, {\mathbf{a}}_{\rm{relax}} \right\}}$, and calculate ${f_{{\rm{UB}}}}\left( {\cal B} \right)$ according to (\ref{f_UB});
		\STATE Update ${\rm{GUB}}$ according to (\ref{GUB});
		\ENDIF
		\STATE Obtain ${\bf{a}}_{\rm{prj}}$, calculate ${f_{{\rm{LB}}}}\left( {\cal B} \right)$ according to (\ref{f_LB});
		\IF {${f_{{\rm{LB}}}}\left( {\cal B} \right) > {\rm{GLB}}$}
		\STATE {${\bf{x}}^* = {\left\{{\mathbf{A}}_{\rm{prj}}, {\mathbf{a}}_{\rm{prj}} \right\}}$, $f^* = f \left( {\bf{a}}_{\rm{prj}}, K_a \right) $;}
		\ENDIF
		\STATE Update ${\rm{GLB}}$ according to (\ref{GLB});
		\STATE {\textit{Pruning:}}
		\STATE Prune ${\cal B}$ if it meets fathomed condition (\ref{Fathomed});
		\STATE Prune non-optimal boxes ${\cal B}' \in {\cal S}$ satisfying (\ref{Nonoptimal});
		\ENDFOR
		\ENDWHILE
	\end{small}
\end{algorithmic}
\end{algorithm}

The complete process of BnB method for the globally optimal PA activation vector is outlined in Algorithm {\ref{alg:BnB}}.

\subsection{Inner Layer (Sub-Optimal Low-Complexity Solution): ISLR-Based PA Activation Vector Optimization}

In this subsection, we develop a low-complexity ISLR-based algorithm for PA activation vector optimization. 
Due to flexible PA activation locations, PAs can be selected that are as close as possible to the UAV, which means lower pathloss. Additionally, due to the power radiation coefficient with proportional attenuation, a smaller subset of PAs should be prioritized. 
Based on the unique characteristics of  flexible activation and proportional power radiation in PASS, we develop this ISLR-based algorithm, where the details are given as follows.

\subsubsection{Channel Gain Calculation and Classification}

The first step in the ISLR algorithm focuses on evaluating and categorizing the PAs based on their channel characteristics. 
Since in-waveguide transmission introduces negligible loss, we mainly consider free-space propagation.
Ignoring time index $t$, the free-space channel gain of $k$-th PA, $h_k$, can be calculated by Eq. (\ref{h_Unk}), which is a key metric that directly reflects the pathloss from the $k$-th PA to UAV.

Using the channel gains, the PAs are divided into two distinct categories to facilitate activation strategies.
PAs ranked in the top $K'$ by descending channel gain $h_k$ are classified as high-efficiency.
Due to their strong signal propagation capabilities, they are given priority in activation as they can contribute more to the received signal with lower power consumption. 
The remaining $K-K'$ PAs, with relatively lower channel gains, are classified as low-efficiency. 
Their activation is treated with caution because their signal contribution per unit of power consumed is lower, and activating them may lead to unnecessary power waste. 
For the convenience of subsequent description, we define the following two sets
\begin{subequations}\label{Classfication}
\begin{align}
& {\mathcal{A}}_{\rm{H}} = \left\{ \pi\left(1\right), \pi\left(2\right), \cdots, \pi\left( K' \right) \right\}, \\
& {\mathcal{A}}_{\rm{L}} = \left\{ \pi\left( K' + 1\right), \pi\left( K' + 2\right), \cdots, \pi\left( K \right) \right\},
\end{align}
\end{subequations}
where $\pi \left(k\right)$ denotes the index of $k$-th activated PA, satisfying $\left|h_{\pi \left(1\right)}\right| \ge \left|h_{\pi \left(2\right)}\right| \ge \cdots \ge \left|h_{\pi \left(K\right)}\right|$.

\subsubsection{Basic Search with Incremental Activation}

Building on the preprocessing step, the basic search phase adopts an incremental approach to explore potential activated PA sets. 
The activation set is initialized as $\mathcal{A}_k$, where $k$ starts from 1 to $K'$, given by
\begin{equation}
\mathcal{A}_k = \left\{ \pi\left(1\right), \pi\left(2\right), \cdots, \pi\left( k \right) \right\}, 1 \le k \le K'.	
\end{equation}
This set represents the first $k$ PAs when considering their channel gains, ensuring that higher-gain PAs are activated earlier to utilize the larger power radiation coefficients. 
Since PAs have been discretely predefined along the waveguide, with given activation set $\mathcal{A}_k$, the arrangement of PAs is determined. 
Thus, the power radiation ratio vector can be calculated according to (\ref{PowerModel}), and the required power consumption $P_{\mathcal{A}_k}$ can be calculated according to (\ref{P_C}). 

Although the pathloss-prioritized basic search phase can initially determine the activation PA set, the complex phase interactions during multi-antenna signal superposition may lead to suboptimal selection, necessitating further refinement. 
To further refine the candidate sets and potentially reduce power consumption while maintaining the communication rate constraint, we employ a local replacement strategy.

\subsubsection{Further Refinement with Adaptive Local Replacement}

Denote the PA activation set obtained from basic search stage by $\mathcal{A}_{k_0}$, and denote the corresponding required power consumption by $P_{\mathcal{A}_{k_0}}$. 
The local replacement phase is extended to allow flexible adjustment of the activation size. This adaptability involves adding a high-gain PA from $\mathcal{A}_{\rm{L}}$ or removing a low-contribution PA from $\mathcal{A}_{k_0}$. The detailed operations are as follows. 

\subsubsubsection{Downsizing Replacement}
For the initial set $\mathcal{A}_{k_0}$, first evaluate each PA $\pi(i) \in \mathcal{A}_{k_0}$ by its marginal gain (MG), which is defined as the reduction in received signal power if the PA is removed, given by
\begin{equation}
	\Delta_i = \sum_{m=1}^{k_0} \left|\beta_m {\tilde{g}}_m h_{\pi(m)} \right| - \sum_{m \neq i} \left| \beta_m {\tilde{g}}_m h_{\pi(m)} \right|.
\end{equation}
PAs with the lowest $\Delta_i$, i.e., minimal contribution to the total signal, are marked as primary replacement targets. 
Remove 1 to $t$ (where $t < k_0$) low-MG PAs from $\mathcal{A}_{k_0}$, forming a smaller set $\mathcal{A}_{k_0 - t}$, constructed as
\begin{equation}
	\mathcal{A}_{k_0-t} = \mathcal{A}_{k_0} \setminus \{\text{top } t \text{ PAs with lowest } \Delta_i\}, t = 1, \cdots k_0-1,
\end{equation}
and recalculate the power radiation ratio vector ${\bm{\beta}}$. 
For each reduced set $\mathcal{A}_{k_0-t}$, calculate the required power consumption $P_{\mathcal{A}_{k_0-t}}$ with (\ref{P_C}). 
If $ \min \limits_{\left\{t = 1, \cdots, k_0-1\right\}} P_{\mathcal{A}_{k_0-t}} < P_{\mathcal{A}_{k_0}}$, retain $\mathcal{A}_{k_0-t}$ with minimum $P_{\mathcal{A}_{k_0-t}}$ as a candidate.

\subsubsubsection{Upsizing Replacement}

Add 1 to $s$ (where $s \leq K - K'$) high-gain PAs from $\mathcal{A}_{\rm{L}}$ (sorted by $|h_{\pi(j)}|$ in descending order) to $\mathcal{A}_{k_0-t}$, forming expanded sets $\mathcal{A}_{k_0 - t + s}$. 
The expanded sets can be constructed as 
\begin{equation}
	\begin{array}{l}
	\mathcal{A}_{k_0 - t +s} = \mathcal{A}_{k_0  - t } \cup \{\pi(K'+1), \cdots, \pi(K'+s)\}, \\
	\;\;\;\;\;\;\;\;\;\; \;\;\;\;\;\;\;\;\;\; \;\;\;\;\;\;\;\;\;\; \;\;\;\;\;\;\;\;\;\; \;\;\;\; s = 1, \cdots, K-K',
	\end{array}
\end{equation}
and recalculate the power radiation ratio vector ${\bm{\beta}}$. 
For each expanded set, calculate the required power consumption $P_{\mathcal{A}_{k_0 - t +s}}$ with (\ref{P_C}). 
If $\min \limits_{\left\{s = 1, \cdots, K-K'\right\}} P_{\mathcal{A}_{k_0 - t+s}} < P_{\mathcal{A}_{k_0 - t}}$, retain $\mathcal{A}_{k_0 - t+s}$ with minimum $P_{\mathcal{A}_{k_0 - t+s}}$ as a candidate, which will be the initial optimization set for next iteration of downsizing replacement.  
Update ${\mathcal{A}}_{\rm{L}}$ with $\left\{ \pi\left( K'+s + 1\right), \pi\left( K' +s+ 2\right), \cdots, \pi\left( K \right) \right\}$.

\subsubsection{Activation Vector Determination}

The procedure of downsizing and upsizing is iteratively repeated. 
To avoid unnecessary computations and ensure efficiency, the algorithm incorporates a pruning mechanism and a final selection step to determine the optimal activation vector.

{\textit{Pruning:}} For downsized sets, the decrease of $k_0$ in is halted if further reducing the size leads to a rise in power consumption, i.e., 
\begin{equation}
P_{\mathcal{A}_{k_0 - t}} \ge P_{\mathcal{A}_{k_0 -(t-1)}}. 
\end{equation} 
The inequality $ \min \limits_{\left\{t = 1, \cdots, k_0-1\right\}} P_{\mathcal{A}_{k_0-t}} < P_{\mathcal{A}_{k_0}}$ can be replaced by the first found $t$ that satisfies $P_{\mathcal{A}_{k_0-t}} < P_{\mathcal{A}_{k_0}}$. 

Similarly, for upsized sets, the increase of $k_0$ in is halted if further expanding the size leads to a rise in power consumption, i.e., 
\begin{equation}
	P_{\mathcal{A}_{k_0 - t+ s}} \ge P_{\mathcal{A}_{k_0 -t + (s-1)}}. 
\end{equation} 
The inequality $\min \limits_{\left\{s = 1, \cdots, K-K'\right\}} P_{\mathcal{A}_{k_0 - t+s}} < P_{\mathcal{A}_{k_0 - t}}$ can be replaced by the first found $s$ that satisfies $ P_{\mathcal{A}_{k_0 - t+s}} < P_{\mathcal{A}_{k_0 - t}}$. 

{\textit{Final selection:}} Among all valid candidate sets identified throughout the base search and refinement phases, the one with the minimum power consumption is selected as the optimal PA activation vector.

\begin{figure}[t]
	\centering
	\includegraphics[width=8.0cm]{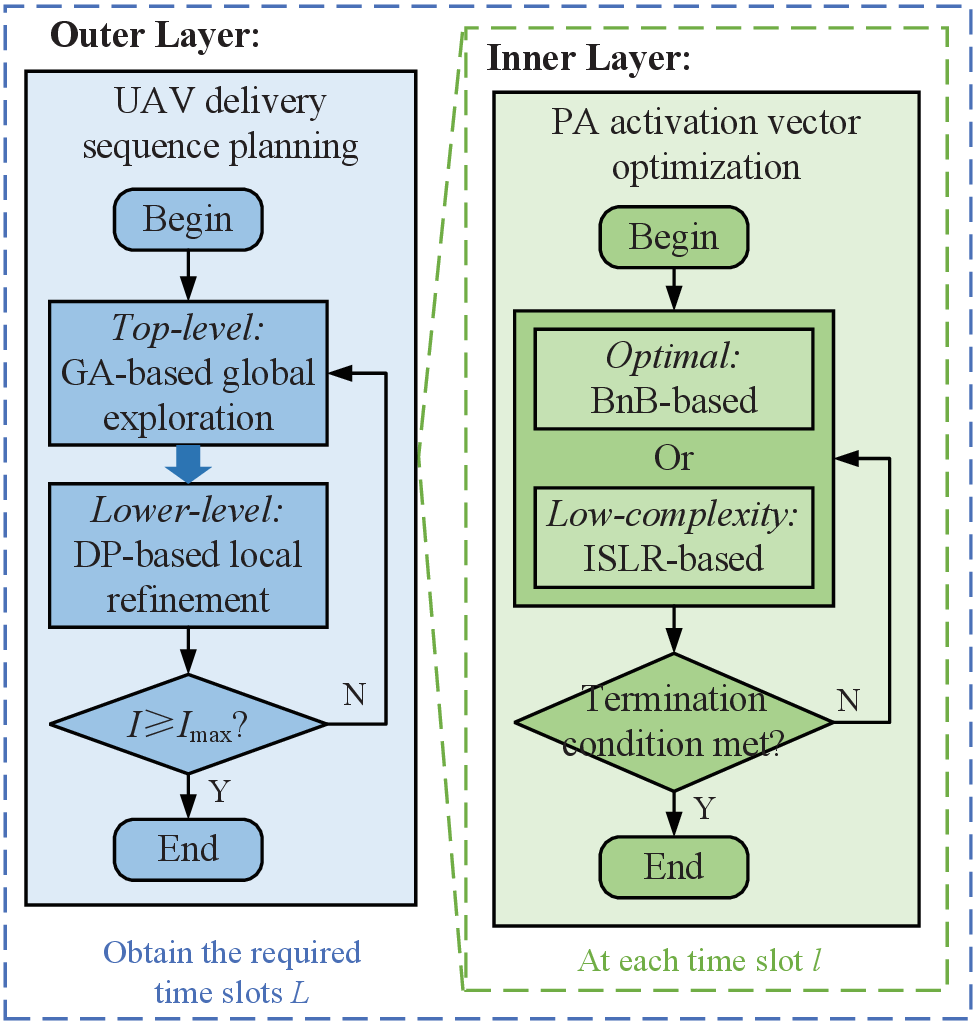}
	\caption{Illustration of the DLO algorithm.}
	\label{AlgorithmChart}
\end{figure}

\begin{algorithm}[t]
	\caption{DLO algorithm for minimizing communication energy consumption}
	\label{alg:Overall}
	\begin{algorithmic}[1]
		\begin{small}
		\REQUIRE 
		Length of each time slot, $\tau$; UAV flight speed, $v_{f}$; Coordinates of delivery nodes; Number of delivery task at each node, $D_{{\rm{task}}, m}$; UAV delivery speed, $v_d$. \\
		\ENSURE
		Optimal energy consumption ${{E}}_{\rm{opt}}$.
		\STATE {\textit{Outer Layer: UAV Delivery Sequence Planning}};
		\STATE Carry out Algorithm {\ref{alg:HAO}}, and obtain the optimal delivering sequence;
		\STATE Calculate the required number of time slots, $L = \sum_{m=1}^{M} \lceil \frac{D_{{\rm{task}}, m}}{v_d \cdot \tau} \rceil + \sum_{m=0}^{M} \lceil \frac{D_{{\rm{path}}, {m \to m+1}}}{v_f \cdot \tau} \rceil$; 
		\STATE {\textit{Inner Layer: PA Activation Vector Optimization}};
		\FOR{$l \le L$}
		\IF{UAV is hovering at $l$- th slot}
		\STATE ${\bf{a}}\left(l\right) = {\bf{a}}\left(l-1\right)$;
		\ELSE
		\STATE {\textit{Case 1: Optimal.}} Carry out Algorithm {\ref{alg:BnB}}, and obtain the PA activation vector ${\bf{a}}\left(l\right)$ at $l$-th time slot;
		\STATE {\textit{Case 2: Low-complexity.}} Carry out the ISLR-base Algorithm, and obtain the PA activation vector ${\bf{a}}\left(l\right)$ at $l$-th time slot;
		\ENDIF
		\STATE Calculate the communication power $P_{\rm{C}} \left(l\right)$ according to (\ref{P_C}), and obtain the energy consumption of communication at $l$-th time slot, $E_{\rm{C}} \left(l\right)= P_{\rm{C}} \left(l\right) \tau$;
		\STATE $l \leftarrow l+1$;
		\ENDFOR
		\STATE Obtain the optimal energy consumption over a flight cycle, ${{E}}_{\rm{opt}} = \sum\limits_{l = 1}^{L} { E_{\rm{C}} \left(l\right)}$.
		\end{small}
	\end{algorithmic}
\end{algorithm}

\subsection{Overall Algorithm}

The proposed DLO algorithm for minimizing communication energy consumption is summarized in Algorithm \ref{alg:Overall}, where the outer-layer UAV delivery sequence planning problem is solved by the HAO scheme, and the inner-layer PA activation optimization problem is solved by the optimal BnB algorithm or the low-complexity ISLR algorithm. 
The illustration of the DLO algorithm is depicted in Fig. \ref{AlgorithmChart}.

The computational complexity of the proposed algorithms is analyzed as follows. The HAO scheme has a complexity determined by both top level and lower level. 
The top-level GA, with population size $N$, maximum evolutionary generations $G_{\text{max}}$, and $M$ delivery nodes, has a complexity of ${\cal{O}}\left(G_{\text{max}} \cdot N \cdot M\right)$ due to operations like selection, crossover, and mutation. 
The lower-level DP, decomposing paths into subpaths each with $a$ nodes, results in a complexity of approximately ${\cal{O}}\left(M \cdot a \cdot 2^a\right)$, leading to an overall HAO complexity of ${\cal{O}}\left(G_{\text{max}} \cdot N \cdot M + M \cdot a \cdot 2^a\right)$.
The BnB algorithm for PA activation vector optimization has a worst-case complexity of ${\cal{O}}\left(2^K\right)$ and a best-case complexity of ${\cal{O}}\left(K\right)$.
The ISLR algorithm has a complexity of ${\cal{O}}\left(K \log K\right)$ due to sorting PAs by channel gain and incremental search with local adjustments.
The total DLO algorithm complexity, combining HAO with either BnB or ISLR, is ${\cal{O}}\left(G_{\text{max}} \cdot N \cdot M + L_f \left( M \cdot a \cdot 2^a + K\right)\right)$ $\sim$ ${\cal{O}}\left(G_{\text{max}} \cdot N \cdot M + L_f \left( M \cdot a \cdot 2^a + 2^K\right)\right)$ when using BnB, and ${\cal{O}}\left(G_{\text{max}} \cdot N \cdot M + L_f \left(M \cdot a \cdot 2^a + K \log K\right)\right)$ when using ISLR, where $ L_f = \sum_{m=0}^{M} \lceil \frac{D_{{\rm{path}}, {m \to m+1}}}{v_f \cdot \tau} \rceil$ denotes the number of flying time slots.

\begin{myRemark}
	{\textit{Although the worst-case complexity of BnB remains exponential, i.e., requiring exhaustive exploration of all possibilities, the actual efficiency highly depends on the tightness of bounds, effectiveness of branching strategies, pruning efficiency and quality of initial solutions. For PASS, the flexibility in PA activation positions directly determines pathloss, which consequently impacts achievable communication rates. Therefore, selecting better initial solutions can significantly enhance the efficiency of the BnB algorithm. }}
\end{myRemark}

\section{Simulation Results} \label{Simulation}

In this section, we provide numerical results to demonstrate the effectiveness of the proposed PASS-UAV design and algorithms. 
The simulation parameters are set as follows.
The system operates at a center frequency of $f_c$ = 15 GHz, with the wavelength of $ \lambda = 0.02$ meters \cite{arXiv:2505.00218}. 
The noise power is set as $\sigma^2$ = -90 dBm. 
Delivery points are randomly distributed in the 3D space of [0, 100] $\times$ [0, 100] $\times$ [2.5, 7.5]. 
The number of tasks at each delivery node is randomly selected from $\left\{1, 2, 3, 4, 5\right\}$. 
The coordinate of the delivery station is (0, 0, 0). 
The waveguide height is 5 meters with stretching
across a span of $L_{\rm{wg}}$ = 100 meters along the $x$-axis direction, and the coordinate of the feed point is (0, 0, 5). 
The number of predefined PAs is set to $K$ = 10.
The PAs are uniformly distributed along the waveguide with a spacing of $L_{\rm{wg}}/K$ meters. 
The refraction index is given by $n_{\rm{eff}}$ = 1.4 \cite{Microwave}.
The constant UAV flight speed is $v_f$ = 5 meters per second, and the delivery speed is $v_d$ = 0.5 tasks per second. 
The required minimum communication rate is set to $R_{\rm{th}}$ = 5 bps/Hz.

The parameters about proposed algorithms are set as follows.
For the GA, the population size is set to 200, with the number of candidate solutions in each generation being 20. 
The probabilities of selection, crossover and mutation are set as $p_{\rm{s}}$ = 0.4, $p_{\rm{c}}$ = 0.6 and $p_{\rm{m}}$ = 0.05, respectively. 
The number of evolutionary generations is $G_{\max}$ = 100.
For the DP algorithm, the length of the subpath is set to $a$ = 3.
The following several benchmark schemes are compared with the proposed scheme.
\begin{itemize}
	\item {\textit{GA:}} GA-based delivery sequence planning. In this benchmark, UAV path planning is accomplished by the GA without DP-based refinement. 
	\item {\textit{MIMO:}} In this benchmark, a conventional MIMO BS is positioned at (0, 0, 5) and equipped with a uniform linear array along $x$-axis comprising $N = 10$ antennas, each connected to a dedicated RF chain. The antenna spacing is set to half the wavelength.
	\item {\textit{FullA:}} Full PA activation scheme. In this benchmark, all predefined PAs are activated. 
\end{itemize}
 
\subsection{UAV Delivery Sequence Planning}

\begin{figure}[t]
	\centering
	\includegraphics[width=7.5cm]{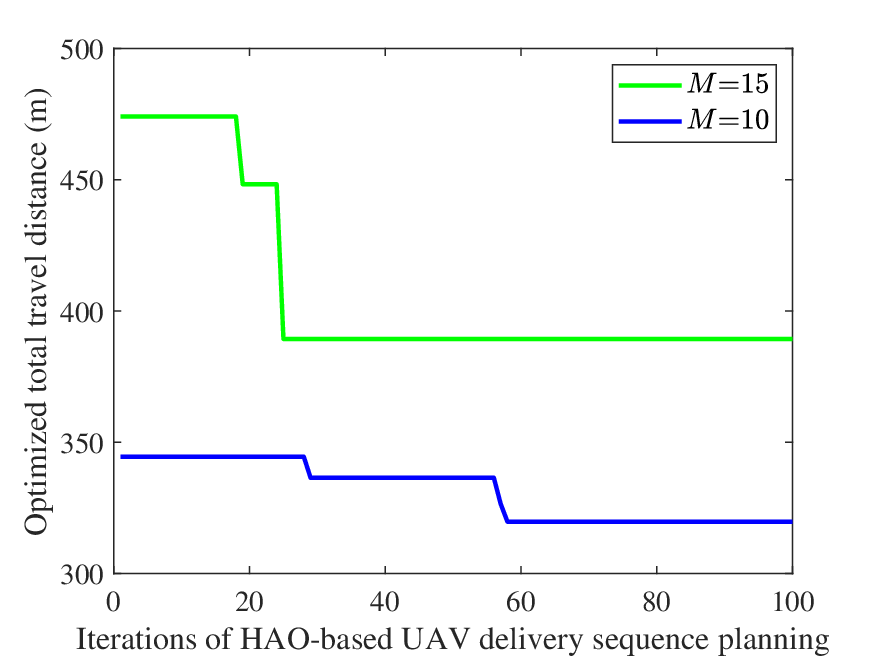}
	\caption{Total flight distance of UAV to complete the delivery tasks.}
	\label{Travel_distance}
\end{figure}

Fig. {\ref{Travel_distance}} illustrates the variation in the total flight distance of the UAV required to complete all delivery tasks with the number of iterations of the HAO scheme. 
It is evident that with the increase in iterations, the total distance exhibits a non-increasing trend and eventually converges to a stable value, which demonstrates the effectiveness of the proposed HAO scheme. 
During the iterative process, there occur instances where the total distance remains unchanged for several consecutive iterations, followed by a subsequent decrease. 
This phenomenon arises because the GA may get trapped in local optimum due to their characteristics of random search and population evolution. 
However, the newly introduced randomly generated population enhances population diversity, thereby facilitating the generation of superior elite solutions. 
Moreover, another obvious phenomenon is that the more delivery nodes, the longer the total flight distance required for the UAV to complete the tasks, which aligns with common sense.

\begin{figure}[t]
	\centering	
	\subfloat[The number of delivery nodes is 10.
	\label{UAV_trajectory_10nodes}]
	{\includegraphics[width=7.5cm]{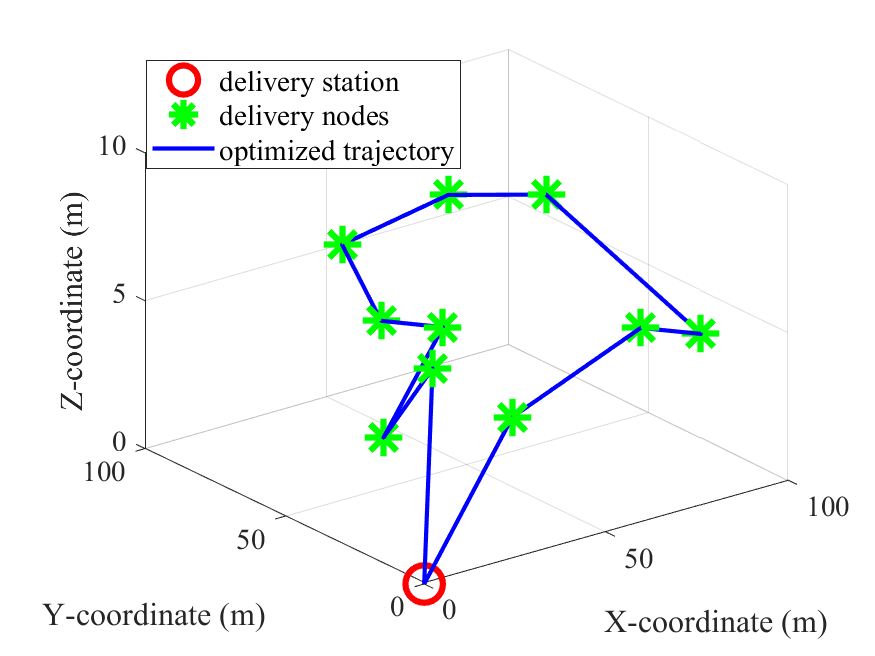}}	
	\vspace{10pt} 	
	\subfloat[The number of delivery nodes is 15.
	\label{UAV_trajectory_15nodes}]
	{\includegraphics[width=7.5cm]{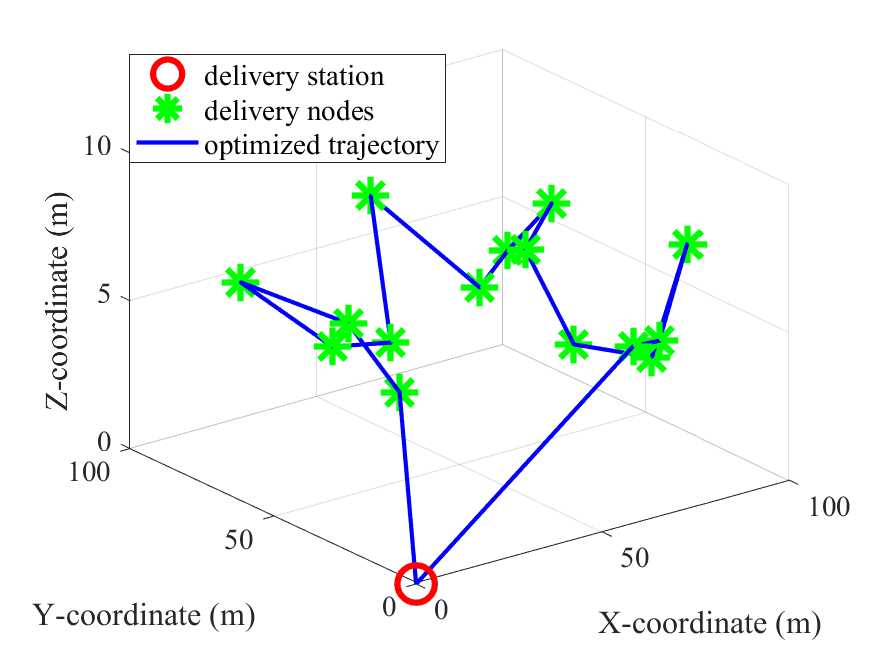}}	
	\caption{Optimized UAV trajectory.}
	\label{DeliveryNodes}
\end{figure}

The optimized UAV trajectory is illustrated in Fig. {\ref{DeliveryNodes}}. It can be observed that the UAV departs from the delivery station and eventually returns to it. 
Moreover, regardless of whether the number of delivery nodes is 10 or 15, the UAV accurately reaches each delivery node, and the trajectory avoids crossing and backtracking, thereby effectively reducing the total path length throughout the entire flight cycle. 
Since we consider a constant UAV flight speed, reducing the flight distance is equivalent to decreasing the flight time required for the UAV to complete all delivery tasks, which will in turn reduce the communication energy consumption within one cycle. 

\subsection{PA Activation Vector Optimization}

\begin{figure}[t]
	\centering
	\includegraphics[width=7.5cm]{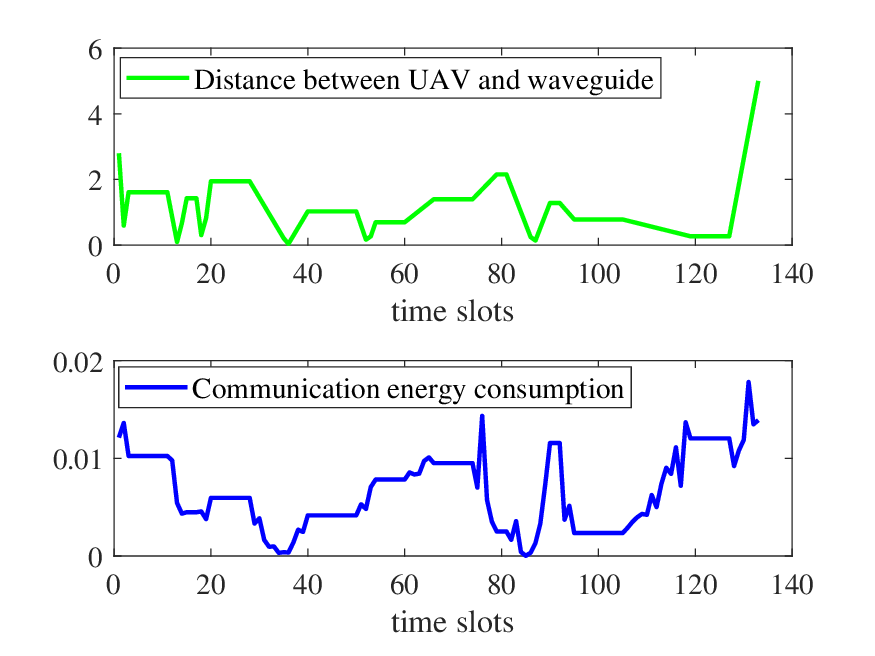}
	\caption{Communication energy consumption versus distance between UAV and waveguide.}
	\label{Distance_Energy}
\end{figure}

The entire UAV flight cycle is divided into multiple time slots. 
In each time slot, the distance between the UAV and the waveguide is calculated with the obtained UAV location. 
BnB algorithm is used to obtain the optimal PA activation vector, thereby obtaining the communication energy consumption in each time slot. 
Fig. {\ref{Distance_Energy}} illustrates how communication energy consumption varies with the distance between the UAV and the waveguide. 
As observed from the figure, energy consumption exhibits a generally positive correlation with distance. Owing to the flexible activation capability of PAs, even though the $x$-coordinate of the UAV varies during flight, the PA closest to it can be activated. 
Consequently, the distance from the UAV to the waveguide is a key determinant of free-space pathloss, i.e., a greater distance leads to more severe pathloss, which in turn requires higher energy consumption to satisfy the communication rate requirement. 
This positive correlation validates the effectiveness of the proposed BnB algorithm. 

\begin{figure}[t]
	\centering
	\includegraphics[width=7.5cm]{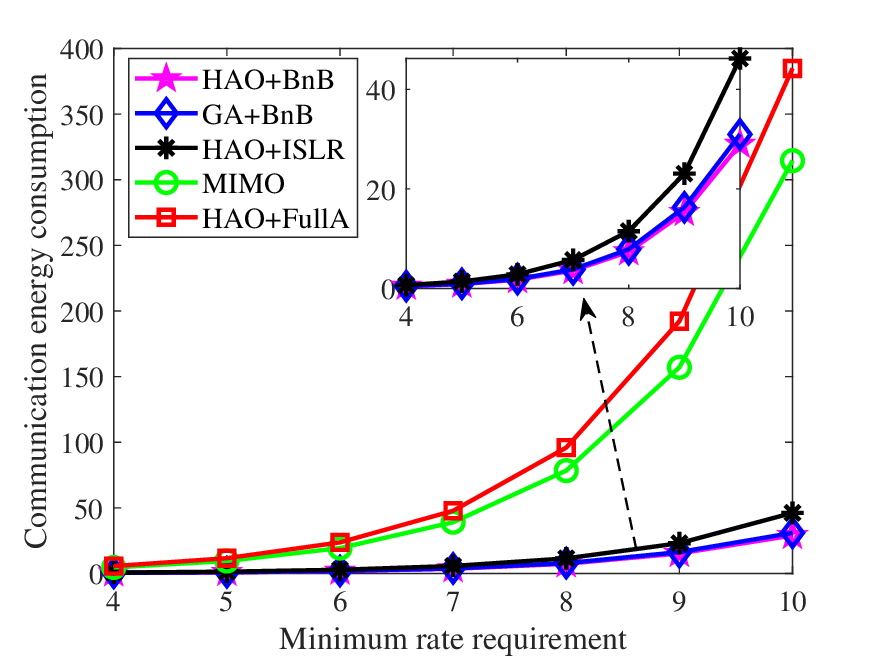}
	\caption{Communication energy consumption versus minimum rate requirement.}
	\label{R_th}
\end{figure}

Fig. \ref{R_th} presents the relationship between communication energy consumption and the minimum rate requirement. 
It can be observed that as the minimum rate requirement increases, the communication energy consumption of all schemes tends to rise. 
Among them, the HAO+BnB, GA+BnB and HAO+ISLR schemes achieve lower energy consumption compared to the MIMO and HAO+FullA schemes. 
This is because the HAO and GA schemes optimize the UAV delivery sequence to reduce flight distance and time, while the BnB and ISLR algorithms efficiently optimize the PA activation vector to minimize communication energy consumption under the rate constraint. 
The FullA scheme exhibits the worst performance because it activates all predefined PAs, yet fails to achieve effective energy concentration due to their discrete distribution, resulting in high communication energy consumption. 
Moreover, the conventional MIMO system shows higher energy consumption compare to the proposed PASS framework with proposed algorithms, especially when the rate requirement is higher, highlighting the superiority of the PASS-enabled framework in handling high-rate communication demands. 
This results confirm that the proposed DLO optimization algorithm effectively reduces communication energy consumption even under increasing rate requirements.

\begin{figure}[t]
	\centering
	\includegraphics[width=7.5cm]{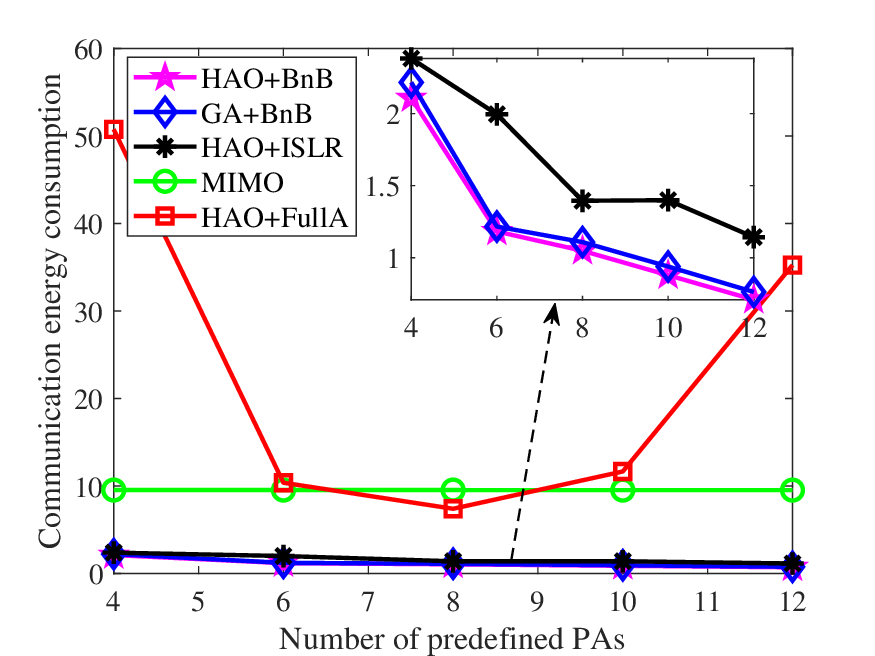}
	\caption{Communication energy consumption versus number of predefined PAs.}
	\label{PA_Num}
\end{figure}

Fig. \ref{PA_Num} shows the relationship between communication energy consumption and the number of predefined PAs. 
Except for FullA, all other schemes exhibit a downward trend in communication energy consumption as the number of predefined PAs increases. 
This is because with more predefined PAs, these schemes can select the most suitable ones (closer to the UAV or with better channel conditions) for activation, thereby reducing pathloss and improving energy efficiency, whereas FullA activates all PAs regardless of quantity, failing to leverage the advantage of more PAs for optimization. 
Moreover, the energy consumption of the FullA scheme shows a trend of first decreasing and then increasing. 
This means that when the full activation mode is adopted, either too few or too many predefined PAs will result in high energy consumption. 
Activating all PAs at preset positions without optimization may even lead to destructive interference due to opposite phases, causing the energy consumption to rise. 
Meanwhile, regardless of changes in the number of predefined PAs, the HAO+BnB, GA+BnB, and HAO+ISLR schemes all achieve good performance. This not only verifies the effectiveness of the proposed algorithms but also confirms the superiority of PASS over traditional MIMO systems.

\section{Conclusions} \label{Conclusion}

A PASS-enabled UAV delivery framework was introduced in this article. 
For minimizing the communication energy consumption in one cycle, a DLO algorithm was developed by jointly optimizing the UAV delivery sequence and the PA activation vector. 
More specifically, a HAO-based scheme was proposed to tackle the NP-hard problem of delivery sequence planning {\textit{at the outer layer}}, and an optimal BnB-based algorithm and a low-complexity ISLR-based algorithm were proposed to resolve this MINLP problem of PA activation vector optimization {\textit{at the inner layer}}. 
The effectiveness of the proposed algorithms was evaluated simulations, which demonstrate that the proposed HAO-based delivery sequence planning scheme effectively reduces total flight distance, while both BnB and ISLR algorithms respectively cut down on communication energy consumption.
Moreover, PASS outperforms conventional MIMO systems, especially with higher required communication rates.

\bibliographystyle{IEEEtran}
\bibliography{ref/ref_PASS_UAV}
	
\end{document}